\documentclass[reprint,superscriptaddress,amsmath,amssymb,aps,pre]{revtex4-2}
\usepackage[utf8]{inputenc}
\usepackage[T1]{fontenc}
\usepackage{graphicx}
\usepackage{mathtools} 
\usepackage{bm}
\usepackage{xcolor}
\usepackage{afterpage}
\usepackage{soul}
\usepackage{dcolumn}
\usepackage{enumitem}
\usepackage[normalem]{ulem}
\usepackage[colorlinks,allcolors=cyan!70!black]{hyperref}

\renewcommand{\selectlanguage}[1]{}

\newcommand{\beq}{\begin{equation}}
\newcommand{\eeq}{\end{equation}}

\renewcommand{\b}{\beta}

\renewcommand{\d}{\delta}
\newcommand{\D}{\Delta}
\newcommand{\e}{\epsilon}
\newcommand{\s}{\sigma}

\newcommand{\bx}{\mathbf{x}} 
\newcommand{\br}{\mathbf{r}} 
\newcommand{\bR}{\mathbf{R}}
\newcommand{\bff}{\mathbf{f}} 
\newcommand{\bv}{\mathbf{v}} 
\newcommand{\be}{\mathbf{e}} 
\newcommand{\bg}{\mathbf{g}} 
\newcommand{\bh}{\mathbf{h}}

\newcommand{\by}{\mathbf{y}}
\newcommand{\bu}{\mathbf{u}}

\newcommand{\bs}{\mathbf{s}}
\newcommand{\half}{\frac{1}{2}}

\newcommand{\mat}[1]{\begin{pmatrix} #1 \end{pmatrix} } 

\let\oldAA\AA
\renewcommand{\AA}{\text{\normalfont\oldAA} }

\begin{document}

\title{Core physical principles of allostery}

\author{Eric Rouviere}
\affiliation{Center for Physics of Evolving Systems, Department of Biochemistry \& Molecular Biology, University of Chicago, Chicago, IL, USA, 60637}
\affiliation{Graduate Program in Biophysical Sciences, University of Chicago, Chicago, IL, USA, 60637}

\author{Olivier Rivoire}
\affiliation{Gulliver, CNRS, ESPCI-Paris, Universit\'e PSL, 75005 Paris, France}

\author{Rama Ranganathan}
\email{ranganathanr@uchicago.edu}
\affiliation{Center for Physics of Evolving Systems, Department of Biochemistry \& Molecular Biology, University of Chicago, Chicago, IL, USA, 60637}
\affiliation{The Pritzker School for Molecular Engineering, University of Chicago, Chicago, IL, USA, 60637}

\begin{abstract}
Allostery is a basic property of proteins that represents the functional coupling of distantly positioned amino acids.  In different manifestations, this property underlies signal transmission, cooperative binding, regulation, catalysis, and evolvability---the elementary features of molecular processes in living cells. While thermodynamic models have provided excellent abstract phenomenological descriptions and classifications of allostery, comparable descriptions and classifications at a more microscopic level, where the nature of physical interactions is specified, are lacking. Starting from minimal assumptions, we derive the essential physical principles of allostery and show that previously described forms of allostery can be viewed as specific implementations of a combination of two core principles: actuation of a soft mode and switching between two states. Organizing allostery by these principles may provide a simpler and more universal language to describe the physics of allostery in proteins.
\end{abstract}


\maketitle

\section*{Introduction}
Allostery is a fundamental property of proteins involved in nearly all biological processes~\cite{monod1963allosteric, may2007allosteric, pellicena2006protein}, representing the elementary step in information flow and regulation in cellular signaling and metabolic networks. Operationally, it is defined by the thermodynamic coupling of different sites of a protein when they do not directly interact, a phenomenological property whose existence is not informative of underlying mechanism. Nevertheless, identifying general principles that support long-range communication within proteins has been a major focus in biophysics to explain how allostery works, to engineer regulation, and to begin understanding its origin in evolution. 

What is the essential problem in defining the general physical principles of allostery? Classical thermodynamic models abstract the high-dimensional conformational space that proteins sample into a few discrete states and define rules for transitioning between these states~\cite{monod1965nature, koshland1966comparison, herzfeld1974general, eigen1968new, eaton2007evolution, hilser2007intrinsic, zhang2018comprehensive}. These models fit experimental data well, produce biochemical parameters that quantify allostery, and have generated valuable conceptual understandings~\cite{monod1965nature, koshland1966comparison,  herzfeld1974general, yifrach1995nested, liu2024parameterized}. An extension is the ensemble allostery model, which generalizes and broadens the application of the classical models by describing allostery as arising from a statistical redistribution of the protein conformational ensemble~\cite{hilser2007intrinsic, hilser2012structural}. This concept can account for global changes in conformational entropy (disorder) as well as concerted structural deformations. But again, these models are independent of mechanism and do not directly address the underlying physics of allostery. In contrast, experimental studies of protein dynamics~\cite{volkman2001two, popovych2006dynamically,hekstra2016electric} and atomistic molecular simulations~\cite{ma2007activation, arora2007large, li2007computational} probe the internal motions of proteins and the forces constraining them. These approaches have provided many detailed case studies of allosteric mechanisms in specific model systems. However, the relationships between these mechanisms are not obvious, leaving open the question of what is universal, if anything, about the physics of allostery. This uncertainty is not particularly surprising; as historically noted for macroscopic machines~\cite{carnot1824reflections}, a detailed characterization of mechanism in individual cases often does not expose core underlying physical principles because it is difficult to separate fundamental rules from idiosyncratic features. In addition, each system only exhibits a portion of the full spectrum of mechanisms available. By focusing on specific cases, general principles can go unnoticed.

One approach to reducing the complexity of allosteric mechanisms is to simply classify by different physical features---multiple conformational states~\cite{monod1965nature, tsai2014unified, ravasio2019mechanics, rivoire2019parsimonious, rouviere2023emergence}, changes in protein dynamics upon binding~\cite{cooper1984allostery, hawkins2004coarse, mcleish2013allostery}, and folding and intrinsic disorder~\cite{hilser2007intrinsic}. But, is this decomposition the right one to identify the unifying physical principles that encompass and underlie all possible mechanisms of allostery? Indeed, what core physical principles are sufficient to produce models for mechanism that are as generally applicable as the thermodynamic models are for phenomenology? In principle, answers to these questions may help reorganize the features noted above in a more physically meaningful manner.

We reasoned that the right approach to these questions is to strip away all the details of specific proteins and consider in the most general way possible the physics by which two spatially distant sites in an abstract model protein can be coupled. The link between the two sites is in the energy landscape of the system, a characteristic defined by the potential function describing the collective pattern of interactions between its constituent parts (i.e., the amino acids). By assuming a generic form of an energy landscape that can accommodate all instances of proteins, we show that only two physical principles can enable allostery. The first is to have ligands engaging the same soft normal modes---the low-energy collective motions of a protein. This type of allostery arises within individual states through ligand-induced changes in either conformation or fluctuations around a conformation, and corresponds to a range of observed mechanisms in both evolved and engineered materials. The second form of allostery involves an energy landscape with two or more distinct macroscopic states that bind ligands with different affinities---the principle that underlies the Monod--Wyman--Changeux (MWC) model~\cite{monod1965nature}. Again, the multi-state form of allostery can involve intramolecular changes in both conformation and fluctuations, and in different limits, corresponds to observed mechanisms in proteins. 

Previous works have proposed unifications of allosteric models at the thermodynamic level~\cite{eigen1968new, herzfeld1974general, hilser2012structural, zhang2018comprehensive}. These descriptions have omitted the notion of soft-mode allostery and its many implementations~\cite{cooper1984allostery, hawkins2004coarse, mcleish2013allostery, zheng2006low, xu2003allosteric, rocks2017designing, yan2017architecture, yan2018principles, dutta2018green}, highlighting the need for the shift in perspective proposed here. The principles derived in this work encompass all known forms of allostery, cast all protein instances as special cases arising from combinations of the core principles, and help shape experimental workflows to classify allosteric mechanisms in new cases. The combination of core physical principles and thermodynamic models for phenomenology should contribute towards a general physical theory of allostery. 

\section*{Materials and Methods}
\noindent \textbf{Conformational Change.} The conformational change is measured as the magnitude of the difference in the mean conformation with both ligand bound, $\langle \br \rangle_{11}$, and no ligand bound $\langle \br \rangle_{00}$,
\beq \label{eq:confchange}
\D\langle\br\rangle = \Vert \langle \br \rangle_{11} - \langle \br \rangle_{00}\Vert.
\eeq
\noindent \textbf{Fluctuation Change.} To measure the change in the magnitude of the fluctuations around the mean conformation, we measure the change in the root-mean-square fluctuation (RMSF) upon binding both ligands,
\beq \label{eq:var}
\begin{split}
\D\mathrm{Var}(\br) = \Big|&\sqrt{\langle \Vert\br\Vert^2 \rangle_{11} - \Vert\langle \br \rangle_{11}\Vert^2} \\
&- \sqrt{\langle \Vert\br\Vert^2 \rangle_{00} - \Vert\langle \br \rangle_{00}\Vert^2}\,\Big|
\end{split}
\eeq
Here, the subscripts on the angle brackets indicate the ligand binding condition for which the average is taken. \\

\noindent \textbf{Conformational Collective Variable.} 
We define a collective variable (CV) $\Phi$ in which the conformational change $\D \bR = \langle \br \rangle_{11} - \langle \br \rangle_{00}$ takes place,
\beq \label{eq:cv_Phi}
\Phi(\br) = \frac{\D \bR}{\Vert \D \bR \Vert} \cdot (\br - \langle \br \rangle_{00}),
\eeq
and use it to define free energy surfaces
\beq\label{eq:freeEnergy_Phi}
\tilde{F}^{\Phi}_{\s_a \s_b}(\phi) = - \frac{1}{\beta}\log \left[ \int d \br \  \d(\Phi(\br)  - \phi)  e^{- \beta U_{\s_a \s_b}(\br)} \right].
\eeq

\noindent \textbf{Disorder Collective Variable.} Transitions such as folding are not well captured by a CV that measures changes along a single conformational direction such as $\Phi$ \eqref{eq:cv_Phi}. Rather, a CV that captures order-disorder transitions is needed, and we propose the following modified RMSF CV,
\beq\label{eq:cv_Psi}
\Psi (\br) = \left\Vert (\br- \langle \br \rangle_{00}) - \left( (\br- \langle \br \rangle_{00}) \cdot \widehat{\D \bR} \right)   \widehat{\D \bR}  \right\Vert,
\eeq
where $\widehat{\D \bR} = \D \bR / \Vert \D \bR \Vert$. This CV measures the conformational deviations in directions other than $\D\bR$. The corresponding free energy surface along $\Psi$ is defined similarly to \eqref{eq:freeEnergy_Phi}. \\

\noindent \textbf{Joint Free Energy Surface.}
A 2d free energy surface in $\Phi$ and $\Psi$ can capture conformational changes and order-disorder changes. 
\beq\label{freeEnergy_PhiPsi}
\begin{split}
\tilde{F}^{\Phi\Psi}_{\s_a \s_b}(\phi, \psi) = &- \frac{1}{\beta}\log \bigg( \frac{1}{h_0^{dN-2}}\int d \br \  \d(\Phi(\br)  - \phi) \\
&\times \d(\Psi(\br) - \psi)  e^{- \beta U_{\s_a \s_b}(\br)} \bigg).
\end{split}
\eeq

\section*{Results}
\subsection*{The model}
\begin{figure}
\centering
\includegraphics[width=0.9\linewidth]{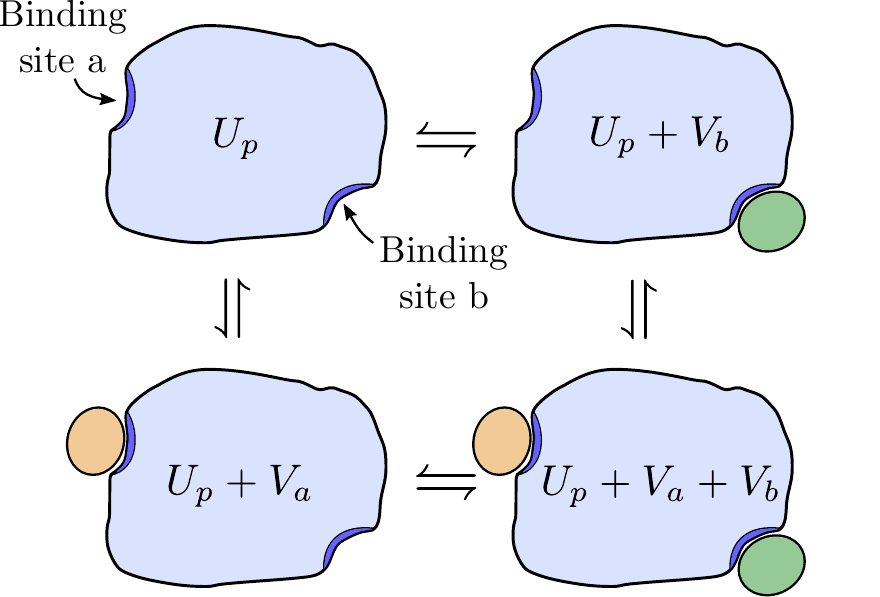}
\caption{The model consists of a protein with two binding sites, $a$ and $b$, resulting in four distinct ligation states: apo, bound at $a$, bound at $b$, and doubly-bound. Ligand binding at sites $a$ and $b$ perturbs the protein's energy landscape, $U_p$, by $V_a$ and $V_b$, respectively.}
\label{fig:formulation_of_model}
\end{figure}

The first step is to formulate a reduced description for a protein that captures only the minimal characteristics to specify the physics of allostery; such a description omits the detail of specific systems and therefore permits the broadest possible conclusions. The ``system'' (blue shape, Fig.~\ref{fig:formulation_of_model}) is comprised of $N$ parts (the amino acids) that interact with each other through a potential $U_p(\br)$ which is a function of the positions of the parts $\br = (\br_1, \br_2, \dots, \br_N)$. The potential $U_p(\br)$ defines the energy landscape of the protein and is defined in more detail below. We then define two binding sites, designated as $a$ and $b$ and marked by orange and green ligands in Fig. \ref{fig:formulation_of_model}.  Ligand binding to these sites causes a change to the energy landscape and is represented mathematically through perturbations $V_a(\br)$ and $V_b(\br)$. For example, the energy landscape of the system when ligand is bound at site $a$ is $U_p(\br)+ V_a(\br)$. Although we describe perturbations in terms of ligand binding, they can equally represent other molecular perturbations, such as post-translational modifications or mutations.  With two sites that can be free, singly-bound, or doubly-bound, there are four ligation conditions of the system. We can write the energy landscapes of these four ligation conditions compactly with indicator variables $\s_a, \s_b$ that take value $1$ if the ligand is bound to the site and $0$ if unbound,  (unbound: $\s_a = \s_b = 0$, bound at $a$: $\s_a = 1, \s_b = 0$, bound at $b$: $\s_a = 0, \s_b = 1$ and doubly-bound: $\s_a = \s_b = 1$) as
\beq \label{eq:FourEnergies}
U_{\s_a \s_b}(\br) = U_p(\br) + \s_a V_a(\br) + \s_b V_b(\br).
\eeq

Proteins are thermal systems that exchange energy with the solvent around them. As a result, their equilibrium properties are determined by the partition function, $Z_{\s_a \s_b} = h_0^{-dN}\int d\br \exp(-\beta U_{\s_a \s_b})$, where $d$ is the number of spatial dimensions. The corresponding free energy for each ligation condition is $F_{\s_a \s_b} = - \beta^{-1} \log Z_{\s_a \s_b}$. Here $\beta = (k_B T)^{-1}$ is the inverse temperature, and we choose units where the length scale $h_0=1$, and it is omitted for clarity going forward.

An allosteric interaction implies that binding at one site alters the binding free energy at the other site, a property known as cooperativity. This feature is quantified as, 
\beq \label{eq:coop}
\D \D F = \left(F_{10} - F_{00} \right)  - \left( F_{11} - F_{01} \right) = -\frac{1}{\beta}\log \left[ \frac{Z_{10}Z_{01}}{Z_{00}Z_{11}} \right].
\eeq
This measures the degree to which the binding free energy of ligand at site $a$ is different in the context of a ligand bound at site $b$. Positive cooperativity occurs when $\D \D F > 0$ and negative cooperativity when $\D \D F < 0$. 

This formulation allows us to connect physical features of the energy landscapes $U_p, V_a, V_b$ to the central thermodynamic property of allostery, $\D\D F$. The basic flow of the work below is to consider how $\D \D F$ and associated properties vary for different forms of the protein energy landscape $U_p(\br)$ and perturbations $V_a(\br)$ and $V_b(\br)$ imposed by ligands.

\subsection*{The two general principles of allostery}
In the simplest case, a ligand acts through a potential $g$ on a single collective motion at each binding site, $\bu_a$ and $\bu_b$, giving
\beq \label{eq:ligandPotentials}
V_a(\br) = g_a(\bu_a^\top \br), \quad V_b(\br) = g_b(\bu_b^\top \br).
\eeq
We show in the Supplementary Information that the results presented below hold for the general case of arbitrary $V_a, V_b$. The partition functions can now be expressed as,
\beq \label{eq:full_Z}
Z_{\s_a \s_b} = \int d x_a  e^{- \s_a \beta g_a(x_a)} \int d x_b e^{- \s_b \beta g_b(x_b)} \ \tilde{Z}(x_a, x_b),
\eeq
\beq \label{eq:eff_partition}
\tilde{Z}(x_a, x_b) = \int d \br   \d(\bu_a^\top \br - x_a)  \d(\bu_b^\top \br - x_b) e^{- \beta U_p(\br)}.
\eeq
The coordinates $x_a,x_b$ specify the binding sites and the procedure in Eqs.~\eqref{eq:full_Z}--\eqref{eq:eff_partition} integrates out the rest of the degrees of freedom $\br$, expressing the system simply as a function of the two binding sites. Taking the log of $\tilde{Z}$ gives the free energy surface of the unperturbed system as a function of the binding site coordinates,
\beq\label{eq:GeneralFreeEnergyLandscape}
\tilde{F}(x_a, x_b) = - \beta^{-1} \log \tilde{Z}(x_a, x_b).
\eeq
With this formulation, we can now discuss the conditions required for cooperativity between the binding sites. If $\tilde{F}$ is separable in terms of the binding sites, $\tilde{F}(x_a, x_b) = \tilde{F}_a(x_a) + \tilde{F}_b(x_b)$, both $\tilde{Z}(x_a, x_b)$ \eqref{eq:eff_partition} and $Z_{\s_a \s_b}$ \eqref{eq:full_Z} factor, meaning no cooperativity ($\D\D F=0$). Thus, cooperativity requires that $\tilde{F}$ be non-separable. A clear implication is that, for two non-local binding sites, the potential for allostery lies in the physical nature of the protein energy landscape $U_p$, independently of the nature of the perturbations at the two sites.

What is the most general representation of $U_p(\br)$ to explore conditions that enable allostery? The energy landscape of a protein may be complex with many minima, each of which defines a macroscopic state. The shape of the energy landscape around each energy minimum $\bR_i$ can be approximated by expanding $U_p(\br)$ to second order,  $U_p(\br) \approx \e_i +  \half (\br-\bR_i)^\top H_i (\br-\bR_i)$, with $\e_i = U_p(\bR_i)$ and $H_i = \nabla \nabla^\top U_p |_{\br=\bR_i}$. $H_i$ is a matrix of second-order derivatives (the ``Hessian'') of $U_p$ evaluated at $\bR_i$. To approximate the full energy landscape, a free energy landscape can be constructed by summing over all such macroscopic states:
\beq \label{eq:multistateLandscape}
F_p( \br) = -\frac{1}{\beta} \log \left( \sum_{i=1}^n e^{-\beta \left[ \half (\br - \bR_i)^\top H_i (\br - \bR_i) + \e_i \right]} \right).
\eeq
This form accurately describes the regions of an energy landscape that contribute most to the free energy---around the energy minima. The corresponding free energy surface is,
\beq \label{eq:multi_state_freeEnergy_surface}
\tilde{F}(x_a,x_b) = -\frac{1}{\beta} \log \left( \sum_{i=1}^n C_i  e^{-\beta \left[ \half \bx_i^\top A_i^{-1} \bx_i + \e_i\right]} \right),
\eeq
where $C_i$ is a state-dependent normalization constant (SI) and,
\beq \label{eq:coarse_grained_allo_matrix}
\bx_i = \mat{x_a - \bu_a^\top \bR_i\\ x_b - \bu_b^\top \bR_i}, \  \mathrm{and} \
A_i = \mat{\bu_a^\top H_i^+ \bu_a & \bu_a^\top H_i^+ \bu_b \\ \bu_a^\top H_i^+ \bu_b & \bu_b^\top H_i^+ \bu_b}.
\eeq
Here, $H_i^+$ represents the pseudo-inverse of $H_i$. The condition of cooperativity is the inseparability of \eqref{eq:multi_state_freeEnergy_surface}. There are two independent ways in which this can occur: 

The first way is a property of the curvature of each basin and is best illustrated when only one basin exists, $n=1$ (for clarity, we omit the subscripts). In this case, the free energy surface, up to an additive constant, simplifies to $\tilde{F}(x_a,x_b) = \half \bx^\top A^{-1}\bx$, and is non-separable only when $\vert \bu_a^\top H^+ \bu_b \vert > 0$. Physically, $\bu_a^\top H^+ \bu_b$ represents the correlation between the collective variables of the two binding sites. Diagonalizing $H^+$ expresses this concept in terms of normal modes, $\bu_a^\top H^+ \bu_b = \bu_a^\top Q  \Lambda^+  Q^\top \bu_b$. The columns of $Q$ are the normal modes of $H$, and $Q^\top \bu_a$ and $Q^\top \bu_b$ describe the overlap of the ligand binding motion with each normal mode. $\Lambda$ is a diagonal matrix whose entries give the stiffnesses of each normal mode.

We can now state the condition of allostery. When $\bu_a^\top H^+ \bu_b$ is small, the cooperativity scales as $\D\D F \sim \bu_a^\top H^+ \bu_b$ (SI). This result implies that when a protein's energy landscape is well approximated by a single quadratic potential, allostery can only occur when the binding sites are coupled through the normal modes. Furthermore, the contribution of a mode is inversely proportional to its stiffness; the magnitude of cooperativity will be maximized when ligands engage the softest modes. We refer to this as the soft mode principle. 

The second way cooperativity can occur is when there are multiple distinct states ($n>1$). To illustrate this alternate route to cooperativity, we consider the simplest case of two states $n=2$, where, in each state, binding sites do not couple through soft normal modes ($\bu_a^\top H_i^+ \bu_b = 0$ for $i=1,2$). Under these conditions, cooperativity is still possible in two cases. First, $\bu_a^\top \bR_1 \neq \bu_a^\top \bR_2$ and $\bu_b^\top \bR_1 \neq \bu_b^\top \bR_2$, corresponding to a situation in which there are two states with different conformations in which the binding sites adopt different shapes. Second, $\bu_a^\top H^+_1 \bu_a \neq \bu_a^\top H^+_2 \bu_a$ and $\bu_b^\top H^+_1 \bu_b \neq \bu_b^\top H^+_2 \bu_b$, corresponding to two states with different curvatures, leading to different entropies and magnitudes of conformational fluctuations at the binding sites. That is, cooperativity occurs when the two states have different conformations or fluctuations that cause ligands to bind to each state with different affinity. This route to cooperativity corresponds to the same principle that underlies the MWC model, and thus we refer to it as the MWC principle.

This calculation shows that regardless of the physical nature of the ligand perturbations $g_a, g_b$, one (or a combination) of two physical principles of the protein's energy landscape must be met for cooperativity to be possible---through soft modes that couple binding sites within a single state (soft mode principle) or by having multiple states with different binding affinities (MWC principle). In the following sections, we consider specific cases and show how these core principles explain known models of allostery.
\begin{figure*}
\centering
\includegraphics[width=0.85\linewidth]{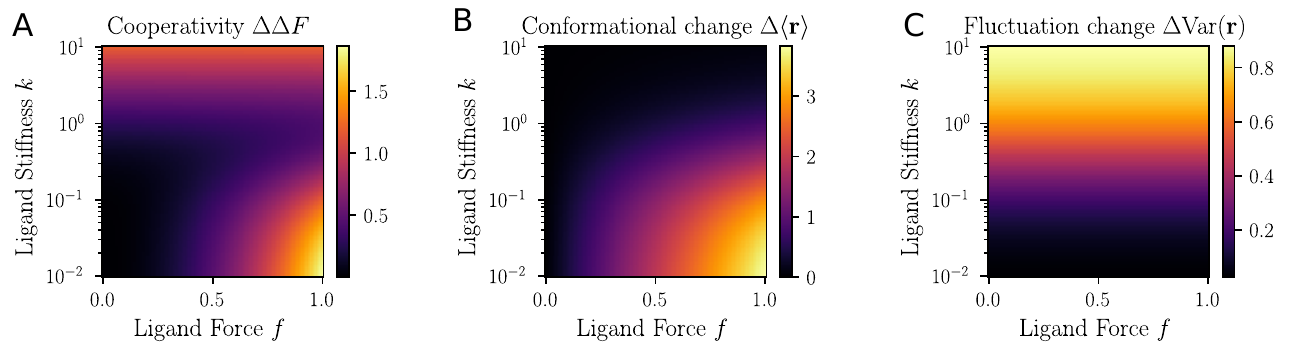}
\caption{Cooperativity with quadratic landscapes. (A) The cooperativity $\D\D F$, (B) the mean conformational change upon ligand binding $\D\langle\br\rangle$ \eqref{eq:confchange}, and (C) the change in the magnitude of fluctuations upon ligand binding $\D\mathrm{Var}(\br)$ \eqref{eq:var}, for a quadratic model with a soft mode, where ligand binding applies a force of magnitude $f$ and a stiffness change of magnitude $k$. The two types of perturbation at each site (force and stiffness) act on the same soft mode that couples the sites. In quadratic landscapes, different types of ligand perturbations can drive cooperativity: applying a force results in a conformational change, while stiffening alters the magnitude of fluctuations. Numerical values for plots are listed in the SI.}
\label{fig:linearModels}
\end{figure*}

\subsection*{Quadratic Models}
What allosteric forms can arise from perturbations within a single macroscopic state? To investigate this, we consider the $n=1$ limit of \eqref{eq:multistateLandscape} with $\bR_1=\mathbf{0}$ and $\e_1=0$, giving the energy landscape, 
\beq\label{eq:Uquad}
U_{\rm quad} = \half \br^\top H \br.
\eeq
To represent ligand potentials that apply forces and fluctuations at binding sites, \eqref{eq:ligandPotentials}, we consider polynomials up to second order, $g_a(x) = \half k_a x^2 - f_a x + c_a$ and $g_b(x) = \half k_b x^2 - f_b x + c_b$. The constant terms do not affect the cooperativity and are set to zero going forward. 

When the ligand perturbations have only first-order terms ($k_a=k_b=0$), they physically correspond to applied forces on the binding sites, $V_a(\br) = -f_a \bu_a^\top \br$ and $V_b(\br) = -f_b \bu_b^\top \br$. In this case, the cooperativity computes to (SI),
\beq \label{eq:coop_force}
\D \D F = f_a f_b \bu_a^\top H^+ \bu_b.
\eeq
\eqref{eq:coop_force} indicates that cooperativity occurs when the ligands apply forces that push on the same soft modes of the protein and push in the same direction. Within the context of a particular physical system, structural rigidity, which is not accounted for in our model, imposes constraints on the normal mode structure. For example, a folded protein must retain stiffness, requiring that only a few soft modes contribute to cooperativity. Negative cooperativity $\Delta \Delta F < 0$ occurs under the same conditions as positive cooperativity, except that ligand forces must push the soft modes in opposite directions. In both cases, this form of cooperativity is purely energetic, with no change in entropy upon ligand binding. This form of cooperativity \eqref{eq:coop_force}, previously calculated in different quadratic allosteric models~\cite{hemery2015evolution, dutta2018green, segers2023mechanisms, balabin2009coarse}, corresponds to the induced-fit case: ligands induce a conformational change by applying forces on binding sites. Each of the energy landscapes has a single minimum, and ligand binding results in the translation of the minimum. 

Recent work has shown that linear elastic networks can be optimized for long-range allosteric tasks~\cite{rocks2017designing, yan2017architecture, yan2018principles, flechsig2017design, dutta2018green, rouviere2023emergence}. These models follow the induced fit case described here---an applied force or displacement to a few parts of the network induces strain along one or several soft modes spanning the network, connecting a distant site. 

Next, we consider second-order ligand perturbations. Setting $f_a = f_b = 0$ yields $V_a(\br) = \tfrac{1}{2} k_a (\bu_a^\top \br)^2$ and $V_b(\br) = \tfrac{1}{2} k_b (\bu_b^\top \br)^2$, which corresponds to stiffening of the binding sites. In proteins, this represents ligand binding events that cause no change in conformation but that dampen coupled protein motions. In this case, the cooperativity computes to
\beq \label{eq:coop_stiff}
\begin{split}
\D \D F &= -\frac{1}{2 \beta} \log \bigg( 1 \\
&- \frac{ k_a k_b \left(\bu_a^\top H^+ \bu_b\right)^2   }
    {(1+ k_a \bu_a^\top H^+ \bu_a)(1+ k_b \bu_b^\top H^+ \bu_b)} \bigg).
\end{split}
\eeq
When both ligands stiffen (or soften) $k_a k_b > 0$, the cooperativity can only be positive and occurs when $\bu_a^\top H^+ \bu_b$ is non-zero. This indicates that soft modes must couple binding sites with soft modes and that ligand binding must stiffen the same soft modes. This case corresponds to the dynamic (or entropic) mechanism proposed by Cooper and Dryden~\cite{cooper1984allostery} and studied extensively in minimal models and elastic networks by McLeish and colleagues~\cite{hawkins2004coarse, mcleish2013allostery}. Additionally, this dynamic case has been considered in a 1D model of DNA allostery~\cite{segers2023mechanisms}. Negative cooperativity can occur when ligand binding stiffens one site but softens the other, $k_a k_b < 0$. Regardless of the sign, the salient point is that this form of cooperativity is purely entropic.

Both the induced fit (force) and dynamic (stiffness) cases operate on the soft mode principle---binding sites are coupled through soft normal modes---the difference lies in the physics of the ligand binding interaction. In the general case, ligand perturbations can both apply forces to and stiffen binding sites. We graphically represent $\D\D F$ for the model under a range of ligand stiffnesses $k = k_a = k_b$ and forces $f = f_a = f_b$. Fig.~\ref{fig:linearModels}A shows two regions of cooperativity, one where force dominates, corresponding to the induced fit case, and the other where stiffness dominates, corresponding to the dynamic case.  

While these two cases operate on the same soft mode principle, physical features differ. This is apparent when representing the change in the mean conformation between the apo (00) and doubly-bound (11) conditions, $\D\langle\br\rangle = \Vert \langle \br \rangle_{11} - \langle \br \rangle_{00}\Vert$ (here the subscript of the angle brackets denotes the ligation condition in which the average is taken). A conformational change occurs in the induced fit regime (Fig.~\ref{fig:linearModels}B). The other regime recapitulates the ``allostery without a conformational change'' feature of dynamic allostery. This regime shows large changes in root mean squared fluctuations $\D\mathrm{Var}(\br)$ \eqref{eq:var} upon binding of both ligands (Fig.~\ref{fig:linearModels}C). 

\subsection*{Bistable Models}
What new forms of cooperativity are possible when the landscape has multiple states? The calculation above (ending with \eqref{eq:multi_state_freeEnergy_surface}) showed that when minima are separated in conformational space ($\bu_a^\top \bR_i \neq \bu_a^\top \bR_j$ and $\bu_b^\top \bR_i \neq \bu_b^\top \bR_j$, for at least one pair of states $i,j$), the free energy surface $\tilde{F}$ is non-separable, allowing for cooperativity. To explore this, we minimally extend the quadratic energy landscape \eqref{eq:Uquad} to introduce two basins in a manner that retains analytical tractability,
\beq \label{eq:bistable_pot}
U_{\rm bi}(\br) = \half \br^\top H \br - |\bg^\top \br|  - \bh^\top \br.
\eeq
Here, bistability is introduced through the absolute value term with minima at $\bR_1 = H^+(\bh + \bg)$ and $\bR_2 = H^+(\bh - \bg)$. Each minimum retains a quadratic basin with curvature $H$, and an additional linear term $\bh^\top \br$ tilts the landscape, enabling one minimum to be lower energy than the other.

When a ligand applies a force ($V_a(\br)=-f_a \bu_a^\top \br$, $V_b(\br)=-f_b \bu_b^\top \br$), the cooperativity takes a lengthy analytical form. However, the essential principles can be illustrated in the zero temperature limit where the free energy converges to the ground state energy ($F \rightarrow E$ as $\beta \rightarrow \infty$). In this case, the ground state cooperativity is,
\beq \label{eq:coop_switch}
\begin{split}
\D\D E &=   f_a f_b \bu_a^\top H^+ \bu_b\\
&+ \left| \left ( f_a \bu_a + f_b \bu_b+ \bh \right)^\top H^+ \bg \right| 
+ \left| \bh ^\top H^{+}  \bg \right| \\
 & - \left| \left (f_a \bu_a + \bh \right)^\top H^+  \bg \right| 
 - \left| \left (f_b \bu_b + \bh \right)^\top H^+ \bg \right|.
\end{split} 
\eeq    
The most important feature is that cooperativity has two components: an induced fit term identical to~\eqref{eq:coop_force} and new terms that depend on the bistability parameter $\bg$. These extra terms describe cases in which the forces $f_a \bu_a$ and $f_b \bu_b$ communicate by coupling to the bistable switch, induced when $\Vert \bg \Vert >0$. Since the cooperativity scales with the bistable parameter, $\D\D E$ can be arbitrarily large, irrespective of the strength of the perturbations, $f_a, f_b$. The contribution from the bistability is maximized when one minimum has initially lower energy, and the ligand forces push the system to stabilize the other minimum, an implementation of the MWC principle. This type of allostery has been called population shift~\cite{tsai2014unified, ravasio2019mechanics}. However, we refer to it as the conformational switch case as it makes clear that there is a physical switch (a transition between two or more states) and that these states differ by their conformation. Negative cooperativity emerges when ligand forces push the system in opposite directions ($(f_a \bu_a^\top H^+ \bg)(f_b \bu_b^\top H^+ \bg) <0$) and is optimal when the two minima are equal energy in the unbound condition (Fig. \ref{fig:neg_coop}).

\begin{figure*}[!t]
\centering
\includegraphics[width=\linewidth]{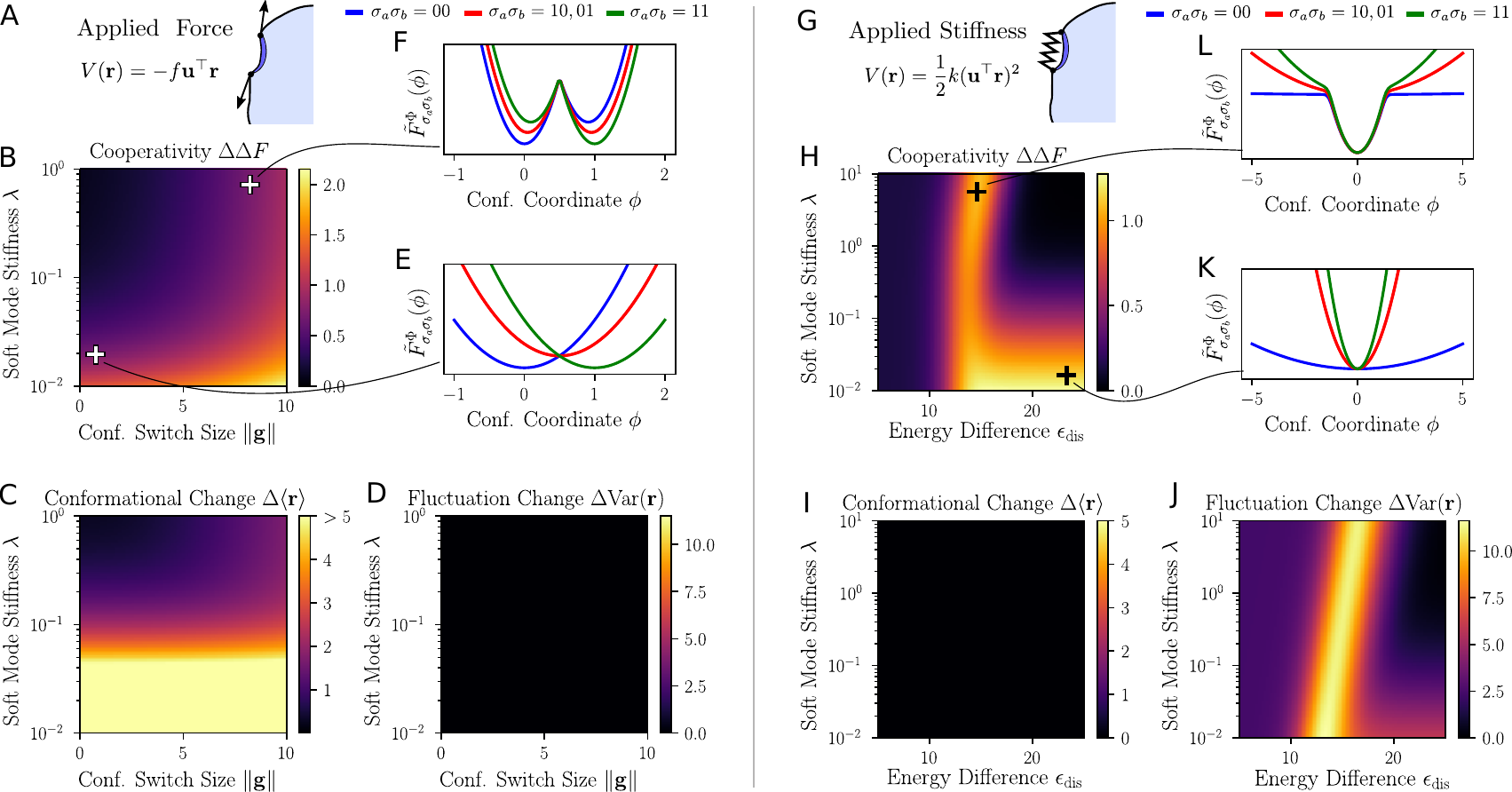}
\caption{Different physical implementations of allosteric principles. The plots on the left half (A–F) describe allosteric cases of a bistable landscape with bistability of strength $\Vert \bg \Vert$ and a mode that couples binding sites with stiffness $\lambda$. Ligand binding is modeled as an applied force (A). 
(B) The cooperativity $\D\D F$, (C) the mean conformational change upon ligand binding $\D\langle\br\rangle$ \eqref{eq:confchange}, and (D) the change in the magnitude of fluctuations upon ligand binding $\D\mathrm{Var}(\br)$ \eqref{eq:var} are plotted over a range of $\Vert \bg \Vert$ and $\lambda$. 
(E) The free energy along the conformational change coordinate, $\tilde{F}^{\Phi}_{\s_a \s_b}(\phi)$, for the system located at the lower-left $+$ mark in (B), for the four ligation conditions. This regime corresponds to the induced fit case.
(F) $\tilde{F}^{\Phi}_{\s_a \s_b}(\phi)$ for the system at the top-right $+$ mark of (B), for the four ligation conditions. This regime corresponds to the conformational switch case.
The plots on the right half (G–L) describe the allosteric cases of a landscape with an ordered state and a disordered state. In the ordered state a mode couples binding sites with stiffness $\lambda$. The two states are separated in energy by $\e_{\rm dis}$, and ligand binding is modeled as an applied stiffness (G).
(H) The cooperativity $\D\D F$, (I) the mean conformational change upon ligand binding $\D\langle\br\rangle$, and (J) the change in the magnitude of fluctuations upon ligand binding $\D\mathrm{Var}(\br)$ are shown over a range of $\lambda$ and $\e_{\rm dis}$.
(K) The free energy along a conformational coordinate, $\tilde{F}^{\Phi}_{\s_a \s_b}(\phi)$, for the system located at the lower-right $+$ mark in (H), for the four ligation conditions. This regime corresponds to the dynamic case. 
(L) $\tilde{F}^{\Phi}_{\s_a \s_b}(\phi)$ for the system at the top-center $+$ mark of (H), for the four ligation conditions.  This regime corresponds to the order-disorder case.
For these landscapes, $U_{\s_a \s_b} = F_{\rm q,d} + \s_a \tfrac{k_a}{2} (\bu_a^\top \br)^2 + \s_b \tfrac{k_b}{2} (\bu_b^\top \br)^2$ in \eqref{eq:freeEnergy_Phi}, and we set $\D \bR$ of \eqref{eq:cv_Phi} 
to be the soft mode, since there is no conformational change upon binding. Numerical values for the plots are listed in the SI.}

\label{fig:mechanisms}
\end{figure*}

In Fig.~\ref{fig:mechanisms}B, we show the cooperativity at nonzero temperature ($\beta=1$) of a system with an internal mode that couples ligand binding sites with stiffness $\lambda$. The conformational switch has magnitude $\Vert \bg \Vert$ under conditions where ligands apply a force. The data show that as in the zero temperature case, cooperativity can be achieved exclusively through the soft mode principle (Fig.~\ref{fig:mechanisms}B, bottom left), the MWC principle by influencing the conformational switch (Fig.~\ref{fig:mechanisms}B, top right), or a combination of the two (Fig.~\ref{fig:mechanisms}B, bottom right). When cooperativity emerges solely from the conformational switch (no soft mode contribution), the system can be analytically written exactly as an MWC model (SI).
Plots of free energy surfaces help to intuitively differentiate various cases. We use the collective variable $\Phi$ defined in \eqref{eq:cv_Phi} that tracks the conformational change and compute free energy surfaces $\tilde{F}^{\Phi}_{\s_a \s_b}(\phi)$
for systems using the induced fit (Fig. \ref{fig:mechanisms}E) and conformational switch (Fig. \ref{fig:mechanisms}F) cases.
Here,  $U_{\s_a \s_b} = U_{\rm bi} - \s_a f_a \bu_a^\top \br - \s_b f_b \bu_b^\top \br$. These computed 1D landscapes recapitulate the landscapes often drawn to illustrate the induced fit and conformational switch cases~\cite{ahuja2019tuning, guo2016protein, wu2024allosteric, tsai2014unified, changeux2011conformational}. Note that both cases in Fig. \ref{fig:mechanisms}E-F are underpinned by a conformational change and have no contribution from changes in conformational fluctuations (compare Figs.~\ref{fig:mechanisms}C-D). An interesting distinction between these two cases is how they scale with the magnitude of the ligand forces. By taking $f_a=f_b=f$, $g = \Vert \bg \Vert$, and $\bh = -f (\bu_a+ \bu_b)/2$, we find that the induced fit component scales as $\kappa^{-1} f^2$ while the conformational switch component scales as $\kappa^{-1} g f$, where $\kappa$ is an effective spring constant. These results suggest that the conformational switch case is preferred when ligand binding imposes a subtle effect (e.g., when the ligand closely fits the geometry of the binding site). 

Many previous physical models used to explain allosteric cooperativity make use of a conformational switch. Examples include elastic networks where the conformational switch is explicitly added~\cite{ravasio2019mechanics}, elastic networks where the conformational switch emerges through an evolutionary selection for cooperativity~\cite{rouviere2023emergence}, and spin-glass models of proteins~\cite{rivoire2019parsimonious}. In addition, many coarse-grained models involve a double (or multi) basin energy landscape imposed by design~\cite{okazaki2006multiple, okazaki2008dynamic, li2014energy,daily2010many,maragakis2005large,chu2007coarse} and display conformational switch-like behavior. 

\subsection*{Order-disorder Models}
Bistable landscapes achieve cooperativity through spatially distinct minima that make the free energy surface $\tilde{F}$ non-separable. However, non-separability can be achieved even when minima overlap ($\bR_i = \bR_j$ for all pairs of states $i,j$) if the magnitude of fluctuations at the binding sites differ between states: $\bu_a^\top H^+_i \bu_a \neq \bu_a^\top H^+_j \bu_a$ and $\bu_b^\top H^+_i \bu_b \neq \bu_b^\top H^+_j \bu_b$ for at least one pair of states $i,j$. This condition occurs in systems with an order-disorder transition where the ordered state is low energy and low entropy, and the disordered state is high energy and high entropy. In proteins, the most obvious implementation of this condition is the global folding transition: the native state is enthalpically stable and lower entropy than the unfolded state~\cite{s1994does}. However, more subtle transitions exist, including partial or local folding~\cite{rundqvist2009noncooperative, schrank2009rational} as well as transitions between a compact yet disordered molten globule and a well-packed native state~\cite{ptitsyn1995molten}. 

To study allostery through order-disorder transitions, we again consider the simplest model that captures this property without loss of generality. Thus, we take a protein energy landscape that comprises a sum of two potentials---an ordered state with a quadratic potential ($U_{\rm quad}$) and another potential to describe the disordered state ($U_{\rm dis}$),
\beq
U_{\rm dis}(\br) = \half \br^\top H_{\rm dis} \br + \e_{\rm dis}.
\eeq
Here $\e_{\rm dis} > 0$ represents the change in energy from breaking interactions between residues in the ordered state. To describe the high conformational entropy of the disordered state in an analytically tractable manner, we consider $\det H_{\rm dis} \ll \det H$, where $\det$ denotes the pseudo-determinant. The two states can then be integrated to produce a free energy landscape,
\beq \label{eq:folding_landscape}
F_{\rm q,d}(\br) = -\beta^{-1} \log \left[
e^{-\beta U_{\rm quad}(\br)} 
+ e^{-\beta U_{\rm dis}(\br)} 
\right].
\eeq

\noindent Since the ordered and disordered states have different entropies, ligand interactions that modulate the entropy have the greatest effect. Thus, we consider the limit where ligand binding just changes the stiffness of the system, $V_a(\br)= \tfrac12 k_a (\bu_a^\top \br)^2$, $V_b(\br)= \tfrac12 k_b (\bu_b^\top \br)^2$. To examine cooperativity, we consider a landscape where the native state curvature $H$ has one internal mode that couples binding sites with stiffness $\lambda$ and an energy difference between ordered and disordered states of $\e_{\rm dis}$. We take the curvature of the disordered state to be isotropic, $H_{\rm dis} \propto I$ such that there is no intrinsic allosteric coupling in this state ($\bu_a^\top H^+_{\rm dis} \bu_b = 0$). 

Fig.~\ref{fig:mechanisms}H shows the cooperativity of this minimal system. At low values of $\e_{\rm dis}$ ($\beta \e_{\rm dis} \ll \tfrac12\log(\det H / \det H_{\rm dis})$) the model is in the disordered state for all ligation conditions and no cooperativity occurs. At high values of $\e_{\rm dis}$ ($\beta \e_{\rm dis} \gg \tfrac12\log(\det H / \det H_{\rm dis})$) the model occupies the ordered state for all ligation conditions and cooperativity occurs through the dynamic case of the soft-mode principle (Fig.~\ref{fig:mechanisms}H bottom right and the associated 1d free energy landscape in Fig.~\ref{fig:mechanisms}K). At intermediate values of $\e_{\rm dis}$, the model is disordered in the absence of ligands and ordered in the presence of ligands. In this case cooperativity occurs even without a soft mode in the ordered state, emerging purely through the switching from disordered to ordered states (Fig.~\ref{fig:mechanisms}H top center and the associated 1d free energy landscape in Fig.~\ref{fig:mechanisms}L). Thus, cooperativity emerges from the relative decrease in entropy of the disordered state upon ligand binding. When cooperativity emerges solely from this feature (no soft mode contribution), the system can again be written exactly as an MWC model (SI). As with the other cases, negative cooperativity in the order–disorder case occurs when ligand binding at different sites has opposing effects: one ligand stiffens the system, while the other increases its flexibility (Fig. \ref{fig:neg_coop}).

This analysis helps to physically unify different forms of allostery described in previous works. In both limits (marked in Fig. \ref{fig:mechanisms}H) we obtain the behavior of ``allostery without conformational change''. However, one case depends on the normal modes within the native state as described by Cooper and Dryden (Fig.~\ref{fig:mechanisms}K)~\cite{cooper1984allostery}, while the other depends on engaging a switch between two states with different entropies (Fig.~\ref{fig:mechanisms}L). Folding-mediated allostery can be seen as a combination of both change in average conformation and overall entropy~\cite{hilser2007intrinsic}.

\begin{figure*}
\centering
\includegraphics[width=0.95\linewidth]{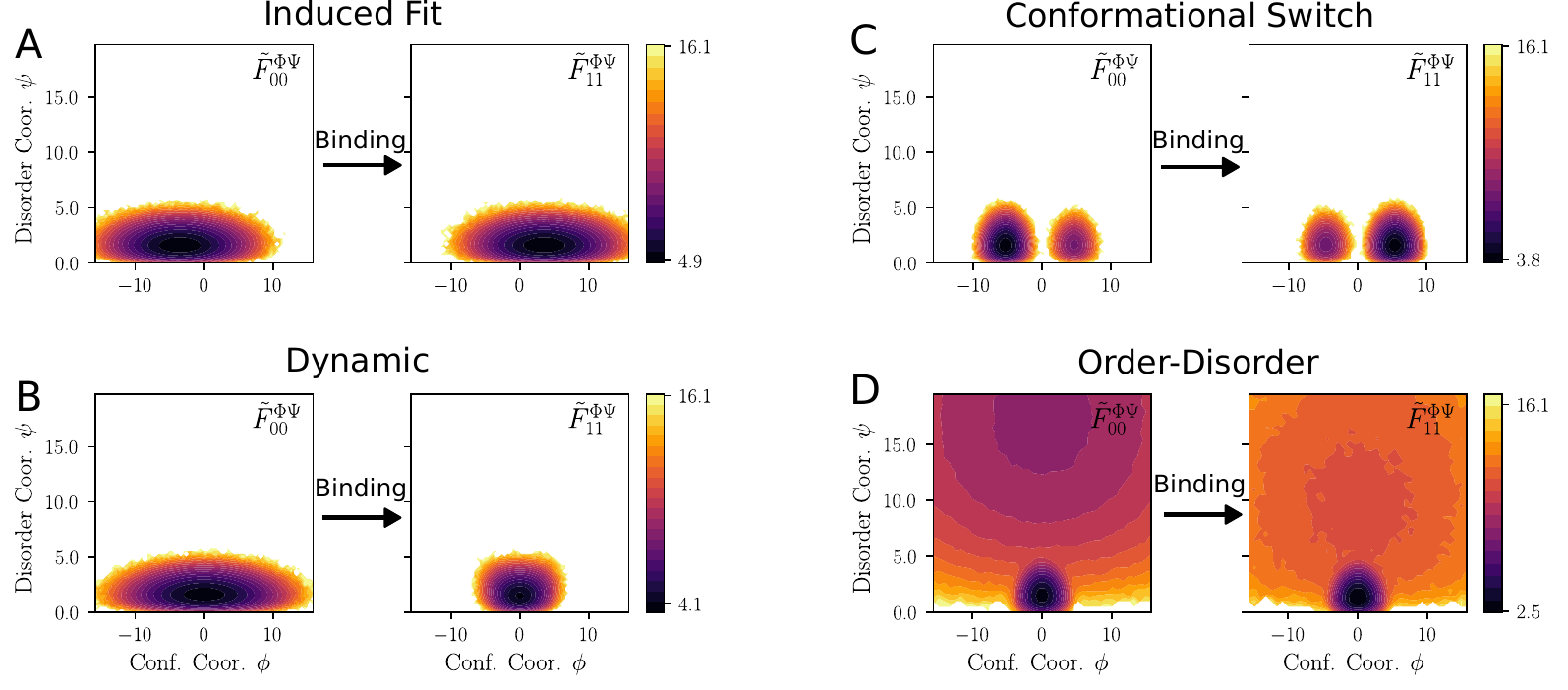}
\caption{
Free energy surfaces along simple collective variables can be used to distinguish different cases of cooperativity. The variable $\phi$ measures progress along a single conformational direction, and the variable $\psi$ measures structural disorder. Each panel shows the free energy surface for the apo and doubly-bound conditions for the (A) induced fit, (B) dynamic, (C) conformational switch, and (D) order-disorder cases. Surfaces are calculated using equilibrium Monte Carlo simulations.}
\label{fig:2d_freeEnergy_landscape}
\end{figure*}

\subsection*{Distinguishing mechanisms in proteins}
The work presented here suggests a different organizational structure for types of allostery than presently used. Current classifications focus on distinguishing whether allostery arises through ligand-induced changes in conformation (the ``structure'' view) or internal disorder (the ``dynamics'' view). The work presented here argues that the central distinguishing characteristic is whether cooperativity operates on the soft mode or the MWC principle, with changes in conformation and dynamics representing an important but orthogonal characteristic. If single-state, allostery must arise through the coupling of ligand perturbations to (soft) normal modes of the protein. If multi-state, allostery may have nothing to do with normal modes, arising instead through biasing the occupancy of different macroscopic states in the free energy landscape. The issue of dynamics is distinct, resolving whether the allosteric mechanism makes use of enthalpic or entropic changes in the system. Consistent with this, the simple models of multi-state allostery making use of either change in conformation (Fig.~\ref{fig:mechanisms}F) or dynamics (Fig.~\ref{fig:mechanisms}L) converge analytically (as expected) to the same thermodynamic model (MWC, SI). Similarly, the requirement for engagement of internal soft modes is true for all forms of single-state allostery, regardless of whether the mechanism involves conformational change or not.

This thinking clarifies the right collective variables (CVs) in which the physics of allostery should be classified. The CV $\Phi$ \eqref{eq:cv_Phi} distinguishes the induced fit (Fig.~\ref{fig:mechanisms}E) and conformational switch cases (Fig.~\ref{fig:mechanisms}F). However, it fails to identify two minima for the two-state order-disorder case because there is no change in the mean conformation (Fig.~\ref{fig:mechanisms}K--L). To address this, we need an orthogonal CV, $\Psi$, \eqref{eq:cv_Psi} that measures root mean squared fluctuations around the conformational direction captured by $\Phi$. As allostery can arise through any combination of these CVs, measuring the free energy surface over both $\Phi$ and $\Psi$, (denoted $\tilde{F}_{\s_a \s_b}^{\Phi \Psi}(\phi, \psi$)) represents a basis to classify allosteric cases. For example, Fig.~\ref{fig:2d_freeEnergy_landscape} shows Monte Carlo estimates of free energy surfaces over these two CVs for the four limiting cases explored analytically in this study. The induced fit (Fig.~\ref{fig:2d_freeEnergy_landscape}A) and dynamic (Fig.~\ref{fig:2d_freeEnergy_landscape}B) cases both operate on the soft mode principle (left panels), and the conformational switch (Fig.~\ref{fig:2d_freeEnergy_landscape}C) and order-disorder (Fig.~\ref{fig:2d_freeEnergy_landscape}D) cases both operate on the MWC principle (right panels). For real proteins, the classification could be achieved experimentally~\cite{hummer2005free, harris2007experimental} or through molecular dynamics simulations~\cite{roux1995calculation, chipot2007free} by plotting free energy landscapes along these collective variables.

\section*{Discussion}

Previous work has stressed the importance of different physical features in allostery: conformational changes, dynamical changes, multiple preexisting conformational states, soft normal modes, intrinsic disorder, and folding. The diversity of these features and the varying levels of abstraction in the models that implement them make it challenging to distinguish fundamental physical principles of allostery from those details that are idiosyncratic to specific systems. In this work, we propose a physical framework in which the essential conditions for allosteric cooperativity can be derived.

The main result is that cooperativity only emerges when the system satisfies one (or a combination) of two physical principles. The first is a soft mode principle where, within a single state, binding sites are coupled through soft normal modes, and ligand binding (or other molecular perturbations) at different sites perturbs the same modes. The second is the MWC principle, which occurs when the system has multiple states with different affinities for the ligands. Upon binding, the system undergoes a switch-like transition between the states. While the MWC principle can be fully defined at a thermodynamic level, namely the MWC model~\cite{monod1965nature}, the soft-mode principle requires a physical description of the energy landscape around a minimum and, by nature, is not captured in thermodynamic models.

We show that each principle can have various specific implementations that recapitulate observed forms of allostery. Both the induced fit and dynamic cases operate on the soft mode principle, with the difference between them arising from how ligand binding perturbs the soft mode---either by applying a force or by stiffening. The conformational switch (or population-shift~\cite{tsai2014unified}) case, where the two states have different conformations, and the order-disorder case, where the two states have different magnitudes of fluctuations, both operate on the MWC principle. The cases we have considered (Fig. \ref{fig:mechanisms}) represent limiting forms of allostery in which we can analytically illustrate and therefore fully understand the mechanism. Specific implementations of allostery in a real protein need not conform strictly to these limiting conditions. However, these more complicated cases can be understood as nothing more than a combination of these archetypal limits. In this sense, we suggest that the core physical principles that distinguish cases of allostery are fully described by those explored in this work.

We have considered a purely equilibrium, thermodynamic definition of allostery as binding cooperativity. As a result, kinetic and non-equilibrium features---including barrier-crossing rates~\cite{graham2005dynamic, robinson2008physical}, the temporal order of molecular events~\cite{hammes2009conformational, okazaki2008dynamic}, and notions of directional signal propagation---are not described. Furthermore, because this work is formulated at the level of energy landscapes, there is no inherent notion of the internal geometry of proteins. As a result, this level of description does not resolve the spatial patterns of amino acid interactions through which forces are transmitted within proteins to mediate coupling between ligand-binding sites. Previous work has explored geometric features of allostery including, how cooperativity should scale with the distance between binding sites and protein size~\cite{yan2018principles,ravasio2019mechanics,segers2023mechanisms}, how structural rigidity contributes to allostery~\cite{thirumalai2019symmetry,yan2018principles}, and how allosteric features originate from the symmetry of oligomeric proteins~\cite{monod1965nature, schulze2014does, thirumalai2019symmetry}. In addition, the sector hypothesis for proteins proposes that allostery is mediated through physically contiguous networks of collectively acting amino acids~\cite{suel2003evolutionarily, halabi2009protein, lee2008surface, reynolds2011hot}. More generally, sectors have been linked to the core constraints that enable native state folding~\cite{socolich2005evolutionary}, active site function~\cite{halabi2009protein, mclaughlin2012spatial, stiffler2015evolvability}, and the capacity of proteins to adapt to fluctuating conditions of selection~\cite{mclaughlin2012spatial, stiffler2015evolvability, raman2016origins}. In this regard, the simple physical models described here represent a foundation for understanding more general rules by which evolution builds allostery in biological macromolecules.

\section*{Author Contributions}
E.R., O.R., and R.R. conceived the project. E.R. performed the calculations. E.R., O.R., and R.R. wrote the article.

\section*{Declaration of interests}
The authors declare no competing interests.

\section*{Acknowledgments}
We thank Kabir Husain, Bryan Andrews, Samar Alqatari, and members of the Ranganathan and Rivoire Labs for helpful discussions. R.R. and O.R. are grateful for funding from the FACCTS program of the France Chicago Center. R.R. acknowledges support from NIH R01GM12345 and R01GM141697, the University of Chicago, and O.R. from ANR-21-CE45-0033 and ANR-22-CE06-0037.

\section*{Data Availability Statement}
The code to generate the figures can be found at \url{https://doi.org/10.5281/zenodo.18689533}.

\bibliography{refs}

@article{monod1963allosteric,
  title={Allosteric proteins and cellular control systems},
  author={Monod, Jacques and Changeux, Jean-Pierre and Jacob, Fran{\c{c}}ois},
  journal={Journal of Molecular Biology},
  volume={6},
  number={4},
  pages={306--329},
  year={1963},
  publisher={Elsevier}
}

@article{may2007allosteric,
  title={Allosteric modulation of {G} protein--coupled receptors},
  author={May, Lauren T and Leach, Katie and Sexton, Patrick M and Christopoulos, Arthur},
  journal={Annu. Rev. Pharmacol. Toxicol.},
  volume={47},
  pages={1--51},
  year={2007},
  publisher={Annual Reviews}
}

@article{pellicena2006protein,
  title={Protein--protein interactions in the allosteric regulation of protein kinases},
  author={Pellicena, Patricia and Kuriyan, John},
  journal={Current Opinion in Structural Biology},
  volume={16},
  number={6},
  pages={702--709},
  year={2006},
  publisher={Elsevier}
}

@article{monod1965nature,
  title={On the nature of allosteric transitions: a plausible model},
  author={Monod, Jacque and Wyman, Jeffries and Changeux, Jean-Pierre},
  journal={J Mol Biol},
  volume={12},
  number={1},
  pages={88--118},
  year={1965}
}

@article{koshland1966comparison,
  title={Comparison of experimental binding data and theoretical models in proteins containing subunits},
  author={Koshland, Jr., DE and Nemethy, G and Filmer, D},
  journal={Biochemistry},
  volume={5},
  number={1},
  pages={365--385},
  year={1966},
  publisher={ACS Publications}
}

@article{hawkins2004coarse,
  title={Coarse-grained model of entropic allostery},
  author={Hawkins, Rhoda J and McLeish, Tom CB},
  journal={Physical Review Letters},
  volume={93},
  number={9},
  pages={098104},
  year={2004},
  publisher={APS}
}

@article{mcleish2013allostery,
  title={Allostery without conformation change: modelling protein dynamics at multiple scales},
  author={McLeish, Thomas CB and Rodgers, TL and Wilson, Mark R},
  journal={Physical Biology},
  volume={10},
  number={5},
  pages={056004},
  year={2013},
  publisher={IOP Publishing}
}

@article{rouviere2023emergence,
  title={Emergence of single-versus multi-state allostery},
  author={Rouviere, Eric and Ranganathan, Rama and Rivoire, Olivier},
  journal={PRX Life},
  volume={1},
  number={2},
  pages={023004},
  year={2023},
  publisher={APS}
}

@article{rivoire2019parsimonious,
  title={Parsimonious evolutionary scenario for the origin of allostery and coevolution patterns in proteins},
  author={Rivoire, Olivier},
  journal={Physical Review E},
  volume={100},
  number={3},
  pages={032411},
  year={2019},
  publisher={APS}
}

@article{rocks2017designing,
  title={Designing allostery-inspired response in mechanical networks},
  author={Rocks, Jason W and Pashine, Nidhi and Bischofberger, Irmgard and Goodrich, Carl P and Liu, Andrea J and Nagel, Sidney R},
  journal={Proceedings of the National Academy of Sciences},
  volume={114},
  number={10},
  pages={2520--2525},
  year={2017},
  publisher={National Acad Sciences}
}

@article{yan2018principles,
  title={Principles for optimal cooperativity in allosteric materials},
  author={Yan, Le and Ravasio, Riccardo and Brito, Carolina and Wyart, Matthieu},
  journal={Biophysical Journal},
  volume={114},
  number={12},
  pages={2787--2798},
  year={2018},
  publisher={Elsevier}
}

@article{ahuja2019tuning,
  title={Tuning the “violin” of protein kinases: The role of dynamics-based allostery},
  author={Ahuja, Lalima G and Taylor, Susan S and Kornev, Alexandr P},
  journal={IUBMB Life},
  volume={71},
  number={6},
  pages={685--696},
  year={2019},
  publisher={Wiley Online Library}
}

@article{ravasio2019mechanics,
  title={Mechanics of allostery: contrasting the induced fit and population shift scenarios},
  author={Ravasio, Riccardo and Flatt, Solange Marie and Yan, Le and Zamuner, Stefano and Brito, Carolina and Wyart, Matthieu},
  journal={Biophysical Journal},
  volume={117},
  number={10},
  pages={1954--1962},
  year={2019},
  publisher={Elsevier}
}

@article{tsai2014unified,
  title={A unified view of “how allostery works”},
  author={Tsai, Chung-Jung and Nussinov, Ruth},
  journal={PLoS Computational Biology},
  volume={10},
  number={2},
  pages={e1003394},
  year={2014},
  publisher={Public Library of Science San Francisco, USA}
}

@article{changeux2011conformational,
  title={Conformational selection or induced fit? 50 years of debate resolved},
  author={Changeux, Jean-Pierre and Edelstein, Stuart},
  journal={F1000 Biology Reports},
  volume={3},
  year={2011},
  publisher={Faculty of 1000 Ltd}
}

@article{eaton2007evolution,
  title={Evolution of allosteric models for hemoglobin},
  author={Eaton, William A and Henry, Eric R and Hofrichter, James and Bettati, Stefano and Viappiani, Cristiano and Mozzarelli, Andrea},
  journal={IUBMB Life},
  volume={59},
  number={8-9},
  pages={586--599},
  year={2007},
  publisher={Wiley Online Library}
}

@article{eigen1968new,
  title={New looks and outlooks on physical enzymology},
  author={Eigen, Manfred},
  journal={Quarterly Reviews of Biophysics},
  volume={1},
  number={1},
  pages={3--33},
  year={1968},
  publisher={Cambridge University Press}
}

@article{herzfeld1974general,
  title={A general approach to co-operativity and its application to the oxygen equilibrium of hemoglobin and its effectors},
  author={Herzfeld, Judith and Stanley, H Eugene},
  journal={Journal of Molecular Biology},
  volume={82},
  number={2},
  pages={231--265},
  year={1974},
  publisher={Elsevier}
}

@article{yifrach1995nested,
  title={Nested cooperativity in the {ATPase} activity of the oligomeric chaperonin {GroEL}},
  author={Yifrach, Ofer and Horovitz, Amnon},
  journal={Biochemistry},
  volume={34},
  number={16},
  pages={5303--5308},
  year={1995},
  publisher={ACS Publications}
}

@article{liu2024parameterized,
  title={A parameterized two-domain thermodynamic model explains diverse mutational effects on protein allostery},
  author={Liu, Zhuang and Gillis, Thomas G and Raman, Srivatsan and Cui, Qiang},
  journal={Elife},
  volume={12},
  pages={RP92262},
  year={2024},
  publisher={eLife Sciences Publications Limited}
}

@article{hilser2007intrinsic,
  title={Intrinsic disorder as a mechanism to optimize allosteric coupling in proteins},
  author={Hilser, Vincent J and Thompson, E Brad},
  journal={Proceedings of the National Academy of Sciences},
  volume={104},
  number={20},
  pages={8311--8315},
  year={2007},
  publisher={National Acad Sciences}
}

@article{cooper1984allostery,
  title={Allostery without conformational change: a plausible model},
  author={Cooper, A and Dryden, DTF},
  journal={European Biophysics Journal},
  volume={11},
  pages={103--109},
  year={1984},
  publisher={Springer}
}

@article{guo2016protein,
  title={Protein allostery and conformational dynamics},
  author={Guo, Jingjing and Zhou, Huan-Xiang},
  journal={Chemical Reviews},
  volume={116},
  number={11},
  pages={6503--6515},
  year={2016},
  publisher={ACS Publications}
}

@article{suel2003evolutionarily,
  title={Evolutionarily conserved networks of residues mediate allosteric communication in proteins},
  author={S{\"u}el, G{\"u}rol M and Lockless, Steve W and Wall, Mark A and Ranganathan, Rama},
  journal={Nature Structural Biology},
  volume={10},
  number={1},
  pages={59--69},
  year={2003},
  publisher={Nature Publishing Group US New York}
}

@article{reynolds2011hot,
  title={Hot spots for allosteric regulation on protein surfaces},
  author={Reynolds, Kimberly A and McLaughlin, Richard N and Ranganathan, Rama},
  journal={Cell},
  volume={147},
  number={7},
  pages={1564--1575},
  year={2011},
  publisher={Elsevier}
}

@article{yan2017architecture,
  title={Architecture and coevolution of allosteric materials},
  author={Yan, Le and Ravasio, Riccardo and Brito, Carolina and Wyart, Matthieu},
  journal={Proceedings of the National Academy of Sciences},
  volume={114},
  number={10},
  pages={2526--2531},
  year={2017},
  publisher={National Acad Sciences}
}

@article{dutta2018green,
  title={Green function of correlated genes in a minimal mechanical model of protein evolution},
  author={Dutta, Sandipan and Eckmann, Jean-Pierre and Libchaber, Albert and Tlusty, Tsvi},
  journal={Proceedings of the National Academy of Sciences},
  volume={115},
  number={20},
  pages={E4559--E4568},
  year={2018},
  publisher={National Acad Sciences}
}

@article{flechsig2017design,
  title={Design of elastic networks with evolutionary optimized long-range communication as mechanical models of allosteric proteins},
  author={Flechsig, Holger},
  journal={Biophysical Journal},
  volume={113},
  number={3},
  pages={558--571},
  year={2017},
  publisher={Elsevier}
}

@article{segers2023mechanisms,
  title={Mechanisms of {DNA}-mediated allostery},
  author={Segers, Midas and Voorspoels, Aderik and Sakaue, Takahiro and Carlon, Enrico},
  journal={Physical Review Letters},
  volume={131},
  number={23},
  pages={238402},
  year={2023},
  publisher={APS}
}

@article{hemery2015evolution,
  title={Evolution of sparsity and modularity in a model of protein allostery},
  author={Hemery, Mathieu and Rivoire, Olivier},
  journal={Physical Review E},
  volume={91},
  number={4},
  pages={042704},
  year={2015},
  publisher={APS}
}

@article{wu2024allosteric,
  title={Allosteric communication and signal transduction in proteins},
  author={Wu, Nan and Barahona, Mauricio and Yaliraki, Sophia N},
  journal={Current Opinion in Structural Biology},
  volume={84},
  pages={102737},
  year={2024},
  publisher={Elsevier}
}

@article{hummer2005free,
  title={Free energy surfaces from single-molecule force spectroscopy},
  author={Hummer, Gerhard and Szabo, Attila},
  journal={Accounts of Chemical Research},
  volume={38},
  number={7},
  pages={504--513},
  year={2005},
  publisher={ACS Publications}
}

@article{harris2007experimental,
  title={Experimental free energy surface reconstruction from single-molecule force spectroscopy using {Jarzynski}’s equality},
  author={Harris, Nolan C and Song, Yang and Kiang, Ching-Hwa},
  journal={Physical Review Letters},
  volume={99},
  number={6},
  pages={068101},
  year={2007},
  publisher={APS}
}

@article{graham2005dynamic,
  title={Dynamic hysteresis in a one-dimensional {Ising} model: Application to allosteric proteins},
  author={Graham, I and Duke, TAJ},
  journal={Physical Review E},
  volume={71},
  number={6},
  pages={061923},
  year={2005},
  publisher={APS}
}

@article{robinson2008physical,
  title={Physical limits on computation by assemblies of allosteric proteins},
  author={Robinson, John M},
  journal={Physical Review Letters},
  volume={101},
  number={17},
  pages={178104},
  year={2008},
  publisher={APS}
}

@article{okazaki2006multiple,
  title={Multiple-basin energy landscapes for large-amplitude conformational motions of proteins: Structure-based molecular dynamics simulations},
  author={Okazaki, Kei-ichi and Koga, Nobuyasu and Takada, Shoji and Onuchic, Jose N and Wolynes, Peter G},
  journal={Proceedings of the National Academy of Sciences},
  volume={103},
  number={32},
  pages={11844--11849},
  year={2006},
  publisher={National Acad Sciences}
}

@article{okazaki2008dynamic,
  title={Dynamic energy landscape view of coupled binding and protein conformational change: induced-fit versus population-shift mechanisms},
  author={Okazaki, Kei-ichi and Takada, Shoji},
  journal={Proceedings of the National Academy of Sciences},
  volume={105},
  number={32},
  pages={11182--11187},
  year={2008},
  publisher={National Acad Sciences}
}

@article{li2014energy,
  title={Energy landscape views for interplays among folding, binding, and allostery of calmodulin domains},
  author={Li, Wenfei and Wang, Wei and Takada, Shoji},
  journal={Proceedings of the National Academy of Sciences},
  volume={111},
  number={29},
  pages={10550--10555},
  year={2014},
  publisher={National Acad Sciences}
}

@article{daily2010many,
  title={Many local motions cooperate to produce the adenylate kinase conformational transition},
  author={Daily, Michael D and Phillips Jr, George N and Cui, Qiang},
  journal={Journal of Molecular Biology},
  volume={400},
  number={3},
  pages={618--631},
  year={2010},
  publisher={Elsevier}
}

@article{maragakis2005large,
  title={Large amplitude conformational change in proteins explored with a plastic network model: adenylate kinase},
  author={Maragakis, Paul and Karplus, Martin},
  journal={Journal of Molecular Biology},
  volume={352},
  number={4},
  pages={807--822},
  year={2005},
  publisher={Elsevier}
}

@article{chu2007coarse,
  title={Coarse-grained free energy functions for studying protein conformational changes: a double-well network model},
  author={Chu, Jhih-Wei and Voth, Gregory A},
  journal={Biophysical Journal},
  volume={93},
  number={11},
  pages={3860--3871},
  year={2007},
  publisher={Elsevier}
}

@article{hammes2009conformational,
  title={Conformational selection or induced fit: a flux description of reaction mechanism},
  author={Hammes, Gordon G and Chang, Yu-Chu and Oas, Terrence G},
  journal={Proceedings of the National Academy of Sciences},
  volume={106},
  number={33},
  pages={13737--13741},
  year={2009},
  publisher={National Acad Sciences}
}

@article{balabin2009coarse,
  title={Coarse-grained modeling of allosteric regulation in protein receptors},
  author={Balabin, Ilya A and Yang, Weitao and Beratan, David N},
  journal={Proceedings of the National Academy of Sciences},
  volume={106},
  number={34},
  pages={14253--14258},
  year={2009},
  publisher={National Acad Sciences}
}

@book{carnot1824reflections,
  title = {Reflections on the Motive Power of Fire and on the Machines Fitted to Develop that Power},
  author = {Carnot, Sadi},
  year = {1824},
  originaltitle = {Réflexions sur la puissance motrice du feu et sur les machines propres à développer cette puissance},
  language = {French},
  publisher = {Chez Bachelier},
  note = {Originally published in 1824.}
}

@article{hilser2012structural,
  title={Structural and energetic basis of allostery},
  author={Hilser, Vincent J and Wrabl, James O and Motlagh, Hesam N},
  journal={Annual Review of Biophysics},
  volume={41},
  number={1},
  pages={585--609},
  year={2012},
  publisher={Annual Reviews}
}

@article{zhang2018comprehensive,
  title={A comprehensive ensemble model for comparing the allosteric effect of ordered and disordered proteins},
  author={Zhang, Luhao and Li, Maodong and Liu, Zhirong},
  journal={PLoS Computational Biology},
  volume={14},
  number={12},
  pages={e1006393},
  year={2018},
  publisher={Public Library of Science San Francisco, CA USA}
}

@article{s1994does,
  title={How does a protein fold?},
  author={Šali, Andrej and Shakhnovich, Eugene and Karplus, Martin},
  journal={Nature},
  volume={369},
  number={6477},
  pages={248--251},
  year={1994},
  publisher={Nature Publishing Group UK London}
}

@article{rundqvist2009noncooperative,
  title={Noncooperative folding of subdomains in adenylate kinase},
  author={Rundqvist, Louise and {\AA}den, Jörgen and Sparrman, Tobias and Wallgren, Marcus and Olsson, Ulrika and Wolf-Watz, Magnus},
  journal={Biochemistry},
  volume={48},
  number={9},
  pages={1911--1927},
  year={2009},
  publisher={ACS Publications}
}

@article{schrank2009rational,
  title={Rational modulation of conformational fluctuations in adenylate kinase reveals a local unfolding mechanism for allostery and functional adaptation in proteins},
  author={Schrank, Travis P and Bolen, D Wayne and Hilser, Vincent J},
  journal={Proceedings of the National Academy of Sciences},
  volume={106},
  number={40},
  pages={16984--16989},
  year={2009},
  publisher={National Acad Sciences}
}

@article{ptitsyn1995molten,
  title={Molten globule and protein folding},
  author={Ptitsyn, OB},
  journal={Advances in Protein Chemistry},
  volume={47},
  pages={83--229},
  year={1995},
  publisher={Elsevier}
}

@article{zheng2006low,
  title={Low-frequency normal modes that describe allosteric transitions in biological nanomachines are robust to sequence variations},
  author={Zheng, Wenjun and Brooks, Bernard R and Thirumalai, D},
  journal={Proceedings of the National Academy of Sciences},
  volume={103},
  number={20},
  pages={7664--7669},
  year={2006},
  publisher={National Acad Sciences}
}

@article{xu2003allosteric,
  title={Allosteric changes in protein structure computed by a simple mechanical model: hemoglobin {T-R2} transition},
  author={Xu, Chunyan and Tobi, Dror and Bahar, I},
  journal={Journal of Molecular Biology},
  volume={333},
  number={1},
  pages={153--168},
  year={2003},
  publisher={Elsevier}
}

@article{ma2007activation,
  title={Activation mechanism of a signaling protein at atomic resolution from advanced computations},
  author={Ma, Liang and Cui, Qiang},
  journal={Journal of the American Chemical Society},
  volume={129},
  number={33},
  pages={10261--10268},
  year={2007},
  publisher={ACS Publications}
}

@article{arora2007large,
  title={Large-scale allosteric conformational transitions of adenylate kinase appear to involve a population-shift mechanism},
  author={Arora, Karunesh and Brooks III, Charles L},
  journal={Proceedings of the National Academy of Sciences},
  volume={104},
  number={47},
  pages={18496--18501},
  year={2007},
  publisher={National Acad Sciences}
}

@article{li2007computational,
  title={A computational investigation of allostery in the catabolite activator protein},
  author={Li, Liwei and Uversky, Vladimir N and Dunker, A Keith and Meroueh, Samy O},
  journal={Journal of the American Chemical Society},
  volume={129},
  number={50},
  pages={15668--15676},
  year={2007},
  publisher={ACS Publications}
}

@article{popovych2006dynamically,
  title={Dynamically driven protein allostery},
  author={Popovych, Nataliya and Sun, Shangjin and Ebright, Richard H and Kalodimos, Charalampos G},
  journal={Nature Structural \& Molecular Biology},
  volume={13},
  number={9},
  pages={831--838},
  year={2006},
  publisher={Nature Publishing Group US New York}
}

@article{thirumalai2019symmetry,
  title={Symmetry, rigidity, and allosteric signaling: from monomeric proteins to molecular machines},
  author={Thirumalai, D and Hyeon, Changbong and Zhuravlev, Pavel I and Lorimer, George H},
  journal={Chemical Reviews},
  volume={119},
  number={12},
  pages={6788--6821},
  year={2019},
  publisher={ACS Publications}
}

@article{volkman2001two,
  title={Two-state allosteric behavior in a single-domain signaling protein},
  author={Volkman, Brian F and Lipson, Doron and Wemmer, David E and Kern, Dorothee},
  journal={Science},
  volume={291},
  number={5512},
  pages={2429--2433},
  year={2001},
  publisher={American Association for the Advancement of Science}
}

@article{hekstra2016electric,
  title={Electric-field-stimulated protein mechanics},
  author={Hekstra, Doeke R and White, K Ian and Socolich, Michael A and Henning, Robert W and {\v{S}}rajer, Vukica and Ranganathan, Rama},
  journal={Nature},
  volume={540},
  number={7633},
  pages={400--405},
  year={2016},
  publisher={Nature Publishing Group UK London}
}

@book{chipot2007free,
  title={Free energy calculations},
  author={Chipot, Christophe and Pohorille, Andrew},
  volume={86},
  year={2007},
  publisher={Springer}
}

@article{roux1995calculation,
  title={The calculation of the potential of mean force using computer simulations},
  author={Roux, Beno{\^\i}t},
  journal={Computer Physics Communications},
  volume={91},
  number={1-3},
  pages={275--282},
  year={1995},
  publisher={Elsevier}
}

@article{halabi2009protein,
  title={Protein sectors: evolutionary units of three-dimensional structure},
  author={Halabi, Najeeb and Rivoire, Olivier and Leibler, Stanislas and Ranganathan, Rama},
  journal={Cell},
  volume={138},
  number={4},
  pages={774--786},
  year={2009},
  publisher={Elsevier}
}

@article{socolich2005evolutionary,
  title={Evolutionary information for specifying a protein fold},
  author={Socolich, Michael and Lockless, Steve W and Russ, William P and Lee, Heather and Gardner, Kevin H and Ranganathan, Rama},
  journal={Nature},
  volume={437},
  number={7058},
  pages={512--518},
  year={2005},
  publisher={Nature Publishing Group UK London}
}

@article{mclaughlin2012spatial,
  title={The spatial architecture of protein function and adaptation},
  author={McLaughlin Jr, Richard N and Poelwijk, Frank J and Raman, Arjun and Gosal, Walraj S and Ranganathan, Rama},
  journal={Nature},
  volume={491},
  number={7422},
  pages={138--142},
  year={2012},
  publisher={Nature Publishing Group UK London}
}

@article{raman2016origins,
  title={Origins of allostery and evolvability in proteins: a case study},
  author={Raman, Arjun S and White, K Ian and Ranganathan, Rama},
  journal={Cell},
  volume={166},
  number={2},
  pages={468--480},
  year={2016},
  publisher={Elsevier}
}

@article{lee2008surface,
  title={Surface sites for engineering allosteric control in proteins},
  author={Lee, Jeeyeon and Natarajan, Madhusudan and Nashine, Vishal C and Socolich, Michael and Vo, Tina and Russ, William P and Benkovic, Stephen J and Ranganathan, Rama},
  journal={Science},
  volume={322},
  number={5900},
  pages={438--442},
  year={2008},
  publisher={American Association for the Advancement of Science}
}

@article{stiffler2015evolvability,
  title={Evolvability as a function of purifying selection in {TEM-1} $\beta$-lactamase},
  author={Stiffler, Michael A and Hekstra, Doeke R and Ranganathan, Rama},
  journal={Cell},
  volume={160},
  number={5},
  pages={882--892},
  year={2015},
  publisher={Elsevier}
}

@article{schulze2014does,
  title={How does symmetry impact the flexibility of proteins?},
  author={Schulze, Bernd and Sljoka, Adnan and Whiteley, Walter},
  journal={Philosophical Transactions of the Royal Society A: Mathematical, Physical and Engineering Sciences},
  volume={372},
  number={2008},
  year={2014},
  publisher={The Royal Society}
}

\clearpage
\onecolumngrid
\section*{Supplementary Information}

\subsection*{Ligand Binding}
Ligand binding is treated as a perturbation of $U_p$ and acts only on the beads that define each binding site. The first binding site, labeled site $a$, is defined by a set of beads $\mathcal{B}_a,$ and the ligand binding perturbation at this site is given by a function $V_a(\br_a)$ where $\br_a = \{ \br_i \}$ for $i \in \mathcal{B}_a$. Similarly, the second binding site, $b$, is comprised of beads in $\mathcal{B}_b$, and the perturbation on this site is given by $V_b(\br_b)$ where $\br_b = \{ \br_i \}$ for $i \in \mathcal{B}_b$.  To describe allosteric cooperativity the sites are taken to have no overlap, $\mathcal{B}_a \cap \mathcal{B}_b=\emptyset$.  To simplify notation in the main text, we write $V_a$ and $V_b$ as functions of all the degrees of freedom $\br$, with it implied that only the binding site degrees of freedom are acted upon.

At each binding site, we defined a linear collective variable (CV): $X_a(\br) = \bu_a^\top \br, X_b(\br) = \bu_b^\top \br$. Here $\bu_a$ and $\bu_b$ are dimensionless unit vectors that have nonzero entries only on beads in $\mathcal{B}_a$ and $\mathcal{B}_b$, respectively, and describe a single collective mode of each binding site. We assume that ligands interact with the protein only along the collective variables: $V_a(\br) = g_a(X_a(\br)), V_b(\br) = g_b(X_b(\br))$.

\subsection*{Multi-basin Free Energy Surface}
Any energy landscape can be well approximated by the following free energy landscape,
\beq
F_p(\br) = -\frac{1}{\beta} \log \sum_{i=1}^n e^{-\beta \left(\half (\br - \bR_i)^\top H_i (\br - \bR_i) + \e_i\right)}.
\eeq
Each state is quadratic with curvature $H_i$, ground state $\bR_i$ and ground state energy $\e_i$. Substituting $F_p$ for $U_p$ in \eqref{eq:eff_partition} and solving the integral, we find,
\begin{align}
\tilde{Z}(&x_a, x_b) = \frac{1}{h_0^{dN-2}} \int d \br  \d(\bu_a^\top \br - x_a)  \d(\bu_b^\top \br - x_b) e^{- \beta F_p(\br)} \\
&= \frac{1}{h_0^{dN-2}}\sum_{i=1}^n \int d \br  \d(\bu_a^\top \br - x_a)  \d(\bu_b^\top \br - x_b) e^{- \beta  \half (\br - \bR_i)^\top H_i (\br - \bR_i) - \beta \e_i } \\
&=\frac{1}{h_0^{dN-2}}\sum_{i=1}^n e^{-\beta \e_i} \int d s_a e^{i2\pi s_a x_a} \int d s_b  e^{i2\pi s_b x_b}  \int d \br \ e^{-\half \beta (\br - \bR_i)^\top H_i (\br - \bR_i) -i 2 \pi  (s_a\bu_a + s_b\bu_b)^\top \br}\\
&= \frac{1}{h_0^{dN-2}} \sum_{i=1}^n e^{-\beta \e_i} \int d s_a e^{i2\pi s_a (x_a - \bu_a^\top \bR_i)} \int d s_b  e^{i2\pi s_b (x_b - \bu_b^\top \bR_i)} \int d \by_i \ e^{-\half \beta \by_i^\top H_i \by_i - i 2 \pi  (s_a\bu_a + s_b\bu_b)^\top \by_i } \\
&=h_0^2\sum_{i=1}^n e^{-\beta \e_i} \sqrt{\frac{(2\pi)^{dN}}{h_0^{2dN} \beta^{dN} \det(H_i)}} 
\int d s_a e^{i2\pi s_a (x_a - \bu_a^\top \bR_i)} \int d s_b  e^{i2\pi s_b (x_b - \bu_b^\top \bR_i)} \left( e^{- 2 \pi^2 \beta^{-1} (s_a\bu_a + s_b\bu_b)^\top H_i^+ (s_a\bu_a + s_b\bu_b) } \right) \\
&= h_0^2 \sum_{i=1}^n e^{-\beta \e_i}  \sqrt{\frac{(2\pi)^{dN}}{ h_0^{2dN}\beta^{dN} \det(H_i)}} 
\int d \bs  e^{-2 \pi^2 \beta^{-1} \bs^\top A_i \bs + i2\pi \bx_i^\top \bs }\\
&= \sum_{i=1}^n  \sqrt{\frac{(2\pi)^{dN-2}}{ h_0^{2(dN-2)} \beta^{dN-2} \det(A_i) \det(H_i)}} 
 e^{-\beta ( \half \bx_i^\top A_i^{-1} \bx_i + \e_i)}
\end{align}
On the third line, we write the delta functions in terms of their Fourier transforms, $\d(t) = \int_{-\infty}^{\infty} e^{-i 2 \pi s t} ds$. On the fourth line we used the substitution $\by_i = \br - \bR_i$ and on the sixth line we introduced $\mathbf{s} = (s_a, s_b)$, $\bx_i = (x_a - \bu_a^\top \bR_i, x_b - \bu_b^\top \bR_i)$ and
\beq
A_i = \mat{\bu_a^\top H_i^+ \bu_a & \bu_a^\top H_i^+ \bu_b \\ \bu_a^\top H_i^+ \bu_b & \bu_b^\top H_i^+ \bu_b}.
\eeq
$d$ is the number of spatial dimensions. The free energy surface is,
\beq \label{eq:freeEnergySurface_SI}
\tilde{F}(x_a,x_b) = -\frac{1}{\beta} \log \left( \sum_{i=1}^n  C_i e^{-\beta ( \half \bx_i^\top A_i^{-1} \bx_i + \e_i)} \right),
\eeq
where, 
\beq
C_i = \sqrt{\frac{(2\pi)^{dN-2}}{ h_0^{2(dN-2)} \beta^{dN-2} \det(A_i) \det(H_i)}}.
\eeq

\subsection*{Conditions for cooperativity}
Cooperativity can only exist when the free energy surface is non-separable, i.e., when $\tilde{F}(x_a, x_b)$ can not be written as $\tilde{F}_a(x_a) + \tilde{F}_b(x_b)$. Since~\eqref{eq:freeEnergySurface_SI} is generally non-separable, we solve the few cases where it is separable to identify which conditions must be broken to achieve non-separability and cooperativity.

\subsubsection*{Single-state case}
When there is only one state $n=1$, \eqref{eq:freeEnergySurface_SI} reduces to 
\beq
\tilde{F}(x_a,x_b) = \half \bx^\top A^{-1}\bx + \e - \frac{1}{\beta} \log (C)
\eeq
which is separable only when the off-diagonal entry of $A^{-1}$ is zero. The subscripts that indicate the state have been dropped. The inverse of a symmetric 2$\times$2 matrix is,
\beq
\mat{a & c \\ c  & b }^{-1} = \frac{1}{ab - c^2} \mat{b & -c \\ -c & a},
\eeq
and so the condition for separability is,
\beq
\frac{-\bu_a^\top H^+ \bu_b}{\bu_a^\top H^+ \bu_a \bu_b^\top H^+ \bu_b - (\bu_a^\top H^+ \bu_b)^2} = 0.
\eeq
Thus in a single state system, cooperativity requires $\bu_a^\top H^+ \bu_b \neq 0$.

\subsubsection*{Multi-state case}
When there are multiple states $n>1$, separability requires $\bu_a^\top H_i^+ \bu_b = 0 \ \forall \ i \in \{1,\dots, n\}$ (assumed in the following) as well as additional conditions. Without loss of generality, we consider the $n=2$ case. The free energy surface~\eqref{eq:freeEnergySurface_SI} reduces to,
\beq
\begin{split}
\tilde{F}(x_a,x_b) = -\frac{1}{\beta} \log  &\Bigg[ C_1 e^{-\beta  \left(\half G^{(1)}_{11} (x_a - \bu_a^\top \bR_1)^2 + \half G^{(1)}_{22} (x_b - \bu_b^\top \bR_1)^2  + \e_1\right)}  \\
&+  C_2 e^{-\beta  \left(\half G^{(2)}_{11} (x_a - \bu_a^\top \bR_2)^2 + \half G^{(2)}_{22} (x_b - \bu_b^\top \bR_2)^2  + \e_2\right)} \Bigg]
\end{split}
\eeq
Here $G^{(i)} = A^{-1}_i$ and $G^{(i)}_{11}$ and $G^{(i)}_{22}$ are the diagonal entries of $G^{(i)}$. $\tilde{F}$ is only separable when the argument of the log takes the form: $ (\mathrm{Function \ of }\ x_a) \times (\mathrm{Function \ of} \ x_b) \times \mathrm{Constant}$. This is the case when either:
\beq
G^{(1)}_{11} = G^{(2)}_{11} \ \mathrm{and} \ \bu_a^\top \bR_1 = \bu_a^\top \bR_2 \quad \mathrm{or}  \quad G^{(1)}_{22} = G^{(2)}_{22} \ \mathrm{and} \ \bu_b^\top \bR_1 = \bu_b^\top \bR_2
\eeq
or both. This implies that non‐separability arises when neither coordinate can be factored out of the log‐sum. In particular, both of the following must hold:
\beq
\bigl(\bu_a^\top H^+_1 \bu_a \neq \bu_a^\top H^+_2 \bu_a \; \mathrm{or} \; \bu_a^\top \bR_1 \neq \bu_a^\top \bR_2\bigr)
\quad\text{and}\quad
\bigl(\bu_b^\top H^+_1 \bu_b \neq \bu_b^\top H^+_2 \bu_b \; \mathrm{or} \; \bu_b^\top \bR_1 \neq \bu_b^\top \bR_2\bigr).
\eeq
In words this means that the two states need to be physically distinct.

For any number of states $n>1$,
\beq
\bigl(\bu_a^\top H^+_i \bu_a \neq \bu_a^\top H^+_j \bu_a \; \mathrm{or} \; \bu_a^\top \bR_i \neq \bu_a^\top \bR_j\bigr)
\quad\text{and}\quad
\bigl(\bu_b^\top H^+_i \bu_b \neq \bu_b^\top H^+_j \bu_b \; \mathrm{or} \; \bu_b^\top \bR_i \neq \bu_b^\top \bR_j\bigr)
\eeq
for at least one pair of states $i,j$.

\subsection*{Scaling of cooperativity within a single basin}
Here we show that, for a single basin ($n=1$), the cooperativity scales as
$\Delta\Delta F \sim \bu_a^\top H^+ \bu_b$ when this quantity is small. For $n=1$, the free energy surface reduces to
\beq
\tilde F(x_a,x_b)=\half \bx^\top A^{-1}\bx + \text{const},
\eeq
where $\bx=(x_a-\bu_a^\top\bR,\;x_b-\bu_b^\top\bR)$ and we define $G=A^{-1}$.
Expanding the quadratic form,
\beq
\tilde F
=\half G_{11}(x_a-\bu_a^\top\bR)^2
+\half G_{22}(x_b-\bu_b^\top\bR)^2
+G_{12}(x_a-\bu_a^\top\bR)(x_b-\bu_b^\top\bR).
\eeq
The partition function for ligation state $(\s_a,\s_b)$ is
\beq
Z_{\s_a\s_b}
=\!\int\!dx_a\,dx_b\,
e^{-\beta[\s_a g_a(x_a)+\s_b g_b(x_b)]}
e^{-\beta \tilde F(x_a,x_b)}.
\eeq
We consider the weak-coupling limit where $\bu_a^\top H^+ \bu_b$ is small.
Since
\beq
G_{12}
=-\frac{\bu_a^\top H^+ \bu_b}
{\bu_a^\top H^+ \bu_a\,\bu_b^\top H^+ \bu_b-(\bu_a^\top H^+\bu_b)^2},
\eeq
this corresponds to small $G_{12}$.
Expanding to first order,
\beq
e^{-\beta G_{12}(x_a-\bu_a^\top\bR)(x_b-\bu_b^\top\bR)}
\simeq
1-\beta G_{12}(x_a-\bu_a^\top\bR)(x_b-\bu_b^\top\bR).
\eeq
Using this expansion,
\beq
Z_{\s_a\s_b}=Z_{\s_a\s_b}^*
-\beta G_{12}\,\Omega_{\s_a\s_b}
+\mathcal{O}(G_{12}^2),
\eeq
where
\begin{align}
Z_{\s_a\s_b}^*
&=\!\int\!dx_a\,e^{-\beta[\s_a g_a(x_a)+\half G_{11}(x_a-\bu_a^\top\bR)^2]}
   \int\!dx_b\,e^{-\beta[\s_b g_b(x_b)+\half G_{22}(x_b-\bu_b^\top\bR)^2]},\\
\Omega_{\s_a\s_b}
&=\!\int\!dx_a\,dx_b\,
e^{-\beta[\s_a g_a(x_a)+\s_b g_b(x_b)]}
e^{-\beta[\half G_{11}(x_a-\bu_a^\top\bR)^2+\half G_{22}(x_b-\bu_b^\top\bR)^2]}
(x_a-\bu_a^\top\bR)(x_b-\bu_b^\top\bR).
\end{align}
Substituting into the cooperativity,
\beq
\frac{Z_{10}Z_{01}}{Z_{00}Z_{11}}
\simeq
1-\beta G_{12}
\frac{\Omega_{10}Z_{01}^*+\Omega_{01}Z_{10}^*
-\Omega_{00}Z_{11}^*-\Omega_{11}Z_{00}^*}
{Z_{00}^*Z_{11}^*},
\eeq
where we used $Z_{10}^*Z_{01}^*=Z_{00}^*Z_{11}^*$ and neglected $\mathcal{O}(G_{12}^2)$ terms.
Finally,
\beq
\Delta\Delta F
=-\beta^{-1}\log\!\left(\frac{Z_{10}Z_{01}}{Z_{00}Z_{11}}\right)
= G_{12}\,\Xi + \mathcal{O}(G_{12}^2),
\eeq
with $\Xi$ a finite, $G_{12}$-independent prefactor.
Thus, for weak coupling,
\beq
\Delta\Delta F \sim G_{12} \sim \bu_a^\top H^+ \bu_b .
\eeq

\subsection*{Derivation of Quadratic Cooperativities \texorpdfstring{\eqref{eq:coop_force} and \eqref{eq:coop_stiff}}{}}
We assume ligand binding affects a single mode \eqref{eq:ligandPotentials}, and consider polynomial perturbations up to second order $g_a(x) = \half k_a x^2 - f_a x$ and $g_b(x) = \half k_b x^2 - f_b x$ (constant terms of the polynomials will not contribute to $\D \D F$ and have been omitted). Since we are considering the single-state case ($n=1$), the subscripts on the parameters are dropped (e.g., $\bx_1 \rightarrow \bx$). We use the notation $A_{ab} = \bu_a^\top H^+ \bu_b$. The full partition function is,
\begin{align}
Z_{\s_a \s_b} &= \int d x_a  e^{- \s_a \beta g_a(x_a)} \int d x_b \ e^{- \s_b \beta g_b(x_b)} e^{-\beta \tilde{F}(x_a, x_b) } 
\end{align}
with,
\beq
\tilde{F}(x_a,x_b) = \half \bx^\top A^{-1}\bx + \e - \frac{1}{\beta} \log (C).
\eeq
Recall, $\bx = (x_a - \bu_a^\top \bR, x_b - \bu_b^\top \bR)$. To compute the cooperativity \eqref{eq:coop} we need to compute the partition function for each binding condition.
\beq
Z_{\s_a \s_b} = C e^{-\beta \e} \int d \bx \; e^{-\beta \left[ \half \bx^\top A^{-1} \bx + \s_a \left(\half k_a x_a^2 - f_a x_a\right) + \s_b \left(\half k_b x_b^2 - f_b x_b\right) \right]}.
\eeq
Which can be rewritten as,
\beq
Z_{\s_a \s_b} = C e^{-\beta \e} \int d \bx  e^{-\beta \left(\half \bx^\top M_{\s_a \s_b} \bx \ - \ \bh_{\s_a \s_b}^\top \bx \ + \ S_{\s_a \s_b} \right)}.
\eeq
where,
\beq
M_{\s_a \s_b} = A^{-1} + \s_a k_a \be_1 \be_1^\top + \s_b k_b \be_2 \be_2^\top,
\eeq
\beq
\bh_{\s_a \s_b} = 
  \s_a\left( f_a - k_a (\bu_a^\top \bR) \right)\be_1 
+ \s_b\left( f_b - k_b (\bu_b^\top \bR) \right)\be_2,
\eeq
and
\beq
S_{\s_a \s_b} = \s_a\left[\half k_a (\bu_a^\top \bR)^2 - f_a\, (\bu_a^\top \bR)\right] + \s_b\left[\half k_b (\bu_b^\top \bR)^2 - f_b\, (\bu_b^\top \bR)\right].
\eeq
Here $\be_1=(1,0), \ \be_2=(0,1)$. This is a Gaussian integral and computes to,
\beq
Z_{\s_a \s_b} = 
\frac{2 \pi}{\beta} C \exp\left(-\beta \e\right) 
\exp\left(-\beta S_{\s_a \s_b} \right)
\exp\left(\frac{\beta}{2} \bh_{\s_a \s_b}^\top M_{\s_a \s_b}^{-1} \bh_{\s_a \s_b}\right)
(\det M_{\s_a \s_b})^{-\tfrac12}.
\eeq
In the ratio of the partition functions in the cooperativity \eqref{eq:coop}, the terms $\frac{2 \pi}{\beta} C \exp\left(-\beta \e\right) 
\exp\left(-\beta S_{\s_a \s_b} \right)$ cancel. To compute the determinants we use the Matrix Determinant Lemma,
\begin{align}
\det(M_{00}) &= \det(A^{-1}) \\
\det(M_{10}) &= \det(A^{-1} + k_a \be_1\be_1^\top ) = \det(A^{-1}) (1 + k_a A_{aa}) \\
\det(M_{01}) &= \det(A^{-1} + k_b \be_2\be_2^\top ) = \det(A^{-1}) (1 + k_b A_{bb}) \\
\det(M_{11}) &= \det(A^{-1} + k_a \be_1\be_1^\top + k_b \be_2\be_2^\top ) = \det(A^{-1}) \left((1 + k_a A_{aa}) (1 + k_b A_{bb}) - k_a k_b A_{ab}^2\right)
\end{align}

We introduce the variables $h_a =  f_a - k_a (\bu_a^\top \bR)$ and $h_b =  f_b - k_b (\bu_b^\top \bR)$ to describe the effective forces that are combinations of the applied forces $f_a,f_b$ and the forces that arise from the applied stiffness. Using the Sherman-Morrison formula,
\begin{align}
\bh_{0 0}^\top M_{00}^{-1} \bh_{00} &= 0 \\  
\bh_{1 0}^\top M_{10}^{-1} \bh_{10} &= \frac{h_a^2 A_{aa}}{1+ k_a A_{aa}} \\  
\bh_{0 1}^\top M_{01}^{-1} \bh_{01} &= \frac{h_b^2 A_{bb}}{1+ k_b A_{bb}} \\ 
\bh_{1 1}^\top M_{11}^{-1} \bh_{11} &= \frac{A_{aa} h_a^2 + A_{bb} h_b^2 + 2 A_{ab} h_a h_b + ( h_a^2 k_b + h_b^2 k_a)(A_{aa}A_{bb} - A_{ab}^2)}{(1 + A_{aa} k_a)(1 + A_{bb} k_b) -  k_a k_b A_{ab}^2}
\end{align}
Putting these together, the full cooperativity becomes
\beq \label{eq:coop_full}
\begin{split} 
\D \D F &= \frac{
2 h_a h_b A_{ab} \left(1 + k_a A_{aa} \right) \left(1 + k_b A_{bb} \right) - A_{ab}^2 \Bigl(k_a h_b^2 (1 + k_a A_{aa}) + k_b h_a^2  (1 + k_b A_{bb}) \Bigr)
}{
2 (1 + k_a A_{aa}) (1 + k_b A_{bb}) \Bigl( (1 + k_a A_{aa}) (1 + k_b A_{bb}) - k_a k_b A_{ab}^2\Bigr)
}  \\[1mm]
&\quad -\frac{1}{2\beta} \log \left( 1 - \frac{k_a k_b A_{ab}^2   }
    {(1+ k_a A_{aa})(1+ k_b A_{bb})}  \right).
\end{split}
\eeq
For the cases of \eqref{eq:coop_force} and \eqref{eq:coop_stiff}, we took $\bR = 0$, leading to $h_a, h_b = f_a, f_b$. When binding purely applies a force, the cooperativity reduces to \eqref{eq:coop_force}.
\beq
\mathrm{When} \ \ k_a=k_b=0 \quad \rightarrow \quad \D \D F =  f_a f_b \bu_a^\top H^+\bu_b,
\eeq
and when binding purely stiffens, with all four ligation conditions mechanically stable ($k_a > -(\bu_a^\top H^+ \bu_a)^{-1}$, $k_b > -(\bu_b^\top H^+ \bu_b)^{-1}$, and $(1+ k_a \bu_a^\top H^+ \bu_a)(1+ k_b \bu_b^\top H^+ \bu_b) > k_a k_b (\bu_a^\top H^+ \bu_b)^2$), the cooperativity reduces to \eqref{eq:coop_stiff},
\beq
\mathrm{When} \ \ f_a=f_b=0 \quad \rightarrow \quad \D \D F =  -\frac{1}{2\beta} \log \bigg( 1 - \frac{ k_a k_b \left(\bu_a^\top H^+ \bu_b\right)^2   }
    {(1+ k_a \bu_a^\top H^+ \bu_a)(1+ k_b \bu_b^\top H^+ \bu_b)} \bigg).
\eeq
Interestingly, when ligand binding increases flexibility at both sites ($- (\bu_a^\top H^+ \bu_a)^{-1} < k_a < 0$ and $- (\bu_b^\top H^+ \bu_b)^{-1} < k_b < 0$), the cooperativity can become arbitrarily large as $k_a$ and $k_b$ approach the mechanical stability bound of the doubly-bound condition, $(1+ k_a \bu_a^\top H^+ \bu_a)(1+ k_b \bu_b^\top H^+ \bu_b) = k_a k_b (\bu_a^\top H^+ \bu_b)^2$, even for an arbitrarily small but nonzero coupling $\bu_a^\top H^+ \bu_b$---a scenario where an optimized soft mode design is not required for significant cooperativity.

Now let us try to simplify \eqref{eq:coop_full}. First we will consider the symmetric case, where $k_a = k_b = k$, $f_a = f_b = f$ and $\bu_a^\top H^+ \bu_a = \bu_b^\top H^+ \bu_b  = \kappa^{-1}$. Next, we nondimensionalize. In this problem, there are three different units: stiffness, force, and energy. The stiffness scale is set by the flexibility of the binding sites, $\kappa$. The relevant energy scale is the thermal energy, $\beta^{-1}$. There is no clear force scale, so it is set by the energy and stiffness scales, $\sqrt{\kappa / \beta}$. Writing the symmetric version of \eqref{eq:coop_full}, 
\beq
\D \D F = \frac{
2 f^2 A_{ab} \Bigl(1 + 2 k \kappa^{-1} + k^2 \kappa^{-2} \Bigr) - A_{ab}^2 \Bigl(2 k f^2 + 2\kappa^{-1} f^2 k^2 \Bigr)
}{
2 (1 + k \kappa^{-1})^2 \Bigl(1 + 2 k \kappa^{-1} + k^2 (\kappa^{-2} - A_{ab}^2)\Bigr)}
-\frac{1}{2\beta} \log \left( 1 - \frac{k^2 A_{ab}^2 }
    {(1+ k \kappa^{-1})^2}  \right).
\eeq
The natural dimensionless variables are 
\beq
\tilde{k} = k / \kappa, \quad \tilde{\eta} = A_{ab} \kappa, \quad \tilde{f} = f / \sqrt{\kappa / \beta}.
\eeq
And,
\beq
\beta \D \D F = \frac{
 \tilde{f}^2 \tilde{\eta} \Bigl(  1 + (1-\tilde{\eta} )\tilde{k}\Bigr) 
}{
 (1 + \tilde{k}) \Bigl( 1 +  2\tilde{k} + (1 - \tilde{\eta} ^2) \tilde{k}^2\Bigr)}
-\frac{1}{2} \log \left( 1 - \tilde{\eta} ^2 \frac{\tilde{k}^2 }
    {(1+ \tilde{k})^2}  \right).
\eeq
In the pure force case $\tilde{k} = 0$, we see $\beta \D\D F = \tilde{\eta} \tilde{f}^2$ and in the pure stiffness case $\tilde{f}=0$, we see $\beta \D\D F = - \tfrac12 \log \left( 1-\tilde{\eta}^2 \tilde{k}^2 / (1 + \tilde{k} )^2 \right)$. 

\subsection*{Derivation of Conformational Switch Cooperativity \texorpdfstring{\eqref{eq:coop_switch}}{}} 

Starting with the bistable energy landscape $U_{\rm bi}(\br)$ defined in \eqref{eq:bistable_pot} and ligand perturbations that apply forces, $V_a(\br) = - f_a \bu_a^\top \br,  \ V_b(\br) = - f_b \bu_b^\top \br$, the energy landscape of each binding condition is,
\beq \label{bistableForceEnergies}
U_{\s_a \s_b}(\br) = U_{\rm bi}(\br) - \s_a f_a \bu_a^\top \br - \s_b f_b \bu_b^\top \br.
\eeq
Each of these energies will have two minima if the following condition is met,
\beq \label{cond42states}
\bg^\top H^+ \left( \bh  + \bff - \bg \right) < 0 < \bg^\top H^+ \left( \bh +  \bff + \bg \right),
\eeq
where $\bff = \s_a f_a \bu_a + \s_b f_b \bu_b$. For a generic quadratic potential with an applied force  $U(\br) = \half \br^\top H\br - \bff^\top \br$ there will be a minimum at $\br_{\rm min} = H^+ \bff$ of energy $E = \min_{\br} U(\br) = -\frac{1}{2} \bff^\top H^+ \bff$. When there are two minima, they will have energies, 
\beq
E_{+} = -\frac{1}{2} (\bff + \bh + \bg)^\top H^+  (\bff + \bh + \bg), \quad \quad E_{-} = -\frac{1}{2} (\bff + \bh - \bg)^\top H^+  (\bff + \bh - \bg)
\eeq
The ground state energy is, 
\begin{align}
E &= \min( E_{+}, E_{-}) = \frac{1}{2} \left( E_{+} + E_{-} - |E_{+} - E_{-}| \right)\\
  &= -\frac{1}{2} \bff^\top H^+ \bff -\frac{1}{2} \bh^\top H^+ \bh - \frac{1}{2} \bg^\top H^+ \bg  - \bff^\top H^+ \bh - | \bff^\top H^+ \bg + \bh^\top H^+ \bg |
\end{align}
The cooperativity of the bistable landscape becomes,
\beq
\begin{split}
    \D \D E = &-\frac{1}{2} f_a^2 \bu_a^\top H^+ \bu_a - \frac{1}{2} \bh^\top H^+ \bh - \frac{1}{2} \bg^\top H^+ \bg - f_a \bu_a^\top H^+ \bh - \left| f_a \bu_a^\top H^+ \bg + \bh^\top H^+ \bg \right| \\
    &- \frac{1}{2} f_b^2 \bu_b^\top H^+ \bu_b - \frac{1}{2} \bh^\top H^+ \bh - \frac{1}{2} \bg^\top H^+ \bg - f_b \bu_b^\top H^+ \bh -  \left| f_b  \bu_b^\top H^+ \bg + \bh^\top H^+ \bg \right| \\
    &+ \frac{1}{2} \bh^\top H^+ \bh + \frac{1}{2} \bg^\top H^+ \bg + \left|\bh^\top H^+ \bg\right| \\
    &+ \frac{1}{2} (f_a \bu_a + f_b \bu_b)^\top H^+ (f_a \bu_a + f_b \bu_b) + \frac{1}{2} \bh^\top H^+ \bh + \frac{1}{2} \bg^\top H^+ \bg + (f_a \bu_a + f_b \bu_b)^\top H^+ \bh \\ 
    &+ \left| (f_a \bu_a + f_b \bu_b)^\top H^+ \bg + \bh^\top H^+ \bg \right| 
\end{split}
\eeq
\beq
\D\D E = f_a f_b \bu_a^\top H^+ \bu_b + \left| \left ( f_a \bu_a + f_b \bu_b+ \bh \right)^\top H^+ \bg \right| + \left| \bh ^\top H^+ \bg \right| - \left| \left (f_a \bu_a + \bh \right)^\top H^+  \bg \right| - \left| \left (f_b \bu_b + \bh \right)^\top H^+ \bg \right|.
\eeq

\clearpage

\subsection*{Connection to MWC model.}
Here we define an MWC model with two binding sites and show that the bistable and order-disorder cases reduce to the MWC model when there is no soft mode contribution to cooperativity.\\

MWC models make two assumptions: there exists an equilibrium of $n>1$ states and, within each state, the binding free energies do not depend on whether a ligand is already bound (i.e., no cooperativity when locked in a single state). For simplicity, we assume the case of a symmetric protein so ligands bind to each site with the same binding energy for a given state.  Denoting the free energy of the unbound condition of the $m$th state as $F^{(m)}_{00}$, the free energy of state $m$ in the $i,j\in\{0,1\}^2$ binding condition can be written as,
\beq
F^{(m)}_{ij} = F_{00}^{(m)} + (i+j) \D F^{(m)}.
\eeq
The cooperativity, as before, is defined as $\D\D F = F_{10} + F_{01} - F_{00} - F_{11}$. The free energy of each ligand binding condition is calculated by summing over states, which in this case we consider $n=2$, $F_{ij} = -\beta^{-1} \log \left( e^{-\beta F_{ij}^{(1)}} + e^{-\beta F_{ij}^{(2)}} \right)$. The cooperativity can be written as, 
\beq
\D\D F = -\beta^{-1} \log \left[ \frac{ 
\left( e^{-\beta F_{10}^{(1)}} + e^{-\beta F_{10}^{(2)}} \right)
\left( e^{-\beta F_{01}^{(1)}} + e^{-\beta F_{01}^{(2)}} \right)
}{ 
\left( e^{-\beta F_{00}^{(1)}} + e^{-\beta F_{00}^{(2)}} \right)
\left( e^{-\beta F_{11}^{(1)}} + e^{-\beta F_{11}^{(2)}} \right)
} \right]
\eeq 
Defining the difference in free energies of the apo states $\D F_{00} = F_{00}^{(2)} - F_{00}^{(1)}$ and the difference of binding energies $\D F = \D F^{(2)} - \D F^{(1)}$ gives the two MWC parameters. With these parameters, the cooperativity becomes,
\beq
\D\D F = -\beta^{-1} \log \left[ \frac{ 
\left( e^{-\beta( F_{00}^{(1)} + \D F^{(1)})}+ e^{-\beta (F_{00}^{(2)} + \D F^{(2)})} \right)^2
}{ 
\left( e^{-\beta F_{00}^{(1)} } + e^{-\beta F_{00}^{(2)} }\right)
\left( e^{-\beta (F_{00}^{(1)} + 2\D F^{(1)})} + e^{-\beta (F_{00}^{(2)} + 2 \D F^{(2)} )} \right)
} \right]
\eeq
The free energies of state 2 can be written in terms of the free energies of state 1 and the differences,
\begin{align}
\D\D F &= -\beta^{-1} \log \left[ \frac{ 
\left( e^{-\beta( F_{00}^{(1)} + \D F^{(1)})}+ e^{-\beta (F_{00}^{(1)} + \D F_{00} + \D F^{(1)} + \D F)} \right)^2
}{ 
\left( e^{-\beta F_{00}^{(1)} } + e^{-\beta (F_{00}^{(1)} + \D F_{00} )}\right)
\left( e^{-\beta (F_{00}^{(1)} + 2\D F^{(1)})} + e^{-\beta (F_{00}^{(1)} + \D F_{00} + 2 \D F^{(1)} + 2\D F )} \right)
} \right]\\
&= -\beta^{-1} \log \left[ \frac{ 
\left( 1 + e^{-\beta (\D F_{00} + \D F)} \right)^2
}{ 
\left( 1 + e^{-\beta \D F_{00} }\right)
\left( 1 + e^{-\beta (\D F_{00} + 2\D F )} \right)
} \right].
\end{align}
The cooperativity is often written in terms of equilibrium constants $c = e^{-\beta \D F}$ and $L = e^{-\beta \D F_{00}}$, 
\beq
\D\D F = -\beta^{-1} \log \left[ \frac{ 
\left( 1 + c L\right)^2
}{ 
\left( 1 + L \right)
\left( 1 + c^2 L \right)
} \right]
\eeq

\noindent \textbf{Order-disorder Model:} The order-disorder model \eqref{eq:folding_landscape} satisfies the two assumptions of the MWC model when $\bu_a^\top H^+ \bu_b=0$. In this case, we can express $\D F$ and $\D F_{00}$ in terms of parameters of the order-disorder landscape \eqref{eq:folding_landscape}. First we solve for $\D F_{00}$. The free energy of the ordered and disordered apo states are,
\beq
F_{00}^{\rm ord} = \frac{1}{2 \beta} \log \left[ \left(  \frac{\beta h_0^2}{2\pi} \right)^{dN} \det H \right], \quad \quad F_{00}^{\rm dis} = \e_{\rm dis} + \frac{1}{2 \beta} \log \left[ \left(  \frac{\beta h_0^2}{2\pi} \right)^{dN} \det H_{\rm dis} \right].
\eeq
Here $\det$ is the pseudo-determinant since $H$ is singular. The difference between these is,
\beq
\D F_{00} =  F_{00}^{\rm dis} -  F_{00}^{\rm ord} =  \e_{\rm dis} + \frac{1}{2 \beta} \log \left[ \frac{\det H_{\rm dis}}{\det H} \right].
\eeq

Next, we solve for $\D F$. The binding free energies in the ordered and disordered states are,
\beq
\begin{split}
\D F^{\rm ord} &= F^{\rm ord}_{10} - F^{\rm ord}_{00} \\
&= \frac{1}{2 \beta} \log \left[ \left(  \frac{\beta h_0^2}{2\pi} \right)^{dN} \det( H + k_a \bu_a \bu_a^\top) \right] - \frac{1}{2 \beta} \log \left[ \left(  \frac{\beta h_0^2}{2\pi} \right)^{dN} \det H \right] = \frac{1}{2 \beta} \log (1 + k_a \bu_a^\top H^+ \bu_a)
\end{split}
\eeq
and,
\beq
\begin{split}
\D F^{\rm dis} &= F^{\rm dis}_{10} - F^{\rm dis}_{00} \\
&= \frac{1}{2 \beta} \log \left[ \left(  \frac{\beta h_0^2}{2\pi} \right)^{dN} \det( H_{\rm dis} + k_a \bu_a \bu_a^\top) \right] - \frac{1}{2 \beta} \log \left[ \left(  \frac{\beta h_0^2}{2\pi} \right)^{dN} \det H_{\rm dis} \right] = \frac{1}{2 \beta} \log (1 + k_a \bu_a^\top H_{\rm dis}^+ \bu_a).
\end{split}
\eeq
Quantities from binding site $b$ do not appear because of symmetry $\bu_a^\top H^+ \bu_a = \bu_b^\top H^+ \bu_b$ and $\bu_a^\top H_{\rm dis}^+ \bu_a = \bu_b^\top H_{\rm dis}^+ \bu_b$. Finally,
\beq
\D F = \D F^{\rm dis} -  \D F^{\rm ord} = \frac{1}{2 \beta} \log \left( \frac{1 + k_a \bu_a^\top H_{\rm dis}^+ \bu_a}{1 + k_a \bu_a^\top H^+ \bu_a} \right).
\eeq

\noindent \textbf{Bistable Model:} The bistable model \eqref{eq:bistable_pot} satisfies the two assumptions of the MWC model when there are two minima: $\bg^\top H^+ \left( \bh  + \bff - \bg \right) < 0 < \bg^\top H^+ \left( \bh +  \bff + \bg \right)$ and no cooperativity exists in each state $\bu_a^\top H^+ \bu_b=0$. Here we express $\D F$ and $\D F_{00}$ in terms of parameters of the bistable landscape. Let state 1 be the basin in $\bg^\top \br > 0$ and state 2 be the basin $\bg^\top \br \leq 0$. We consider conditions where the basins are distinct (i.e., the barrier is much larger than the thermal energy $\beta^{-1}$). First we solve for $\D F_{00}$.
\beq
F^{(1)}_{00}= \frac{1}{2 \beta} \log \left[ \left(  \frac{\beta h_0^2}{2\pi} \right)^{dN} \det H \right] - \half (\bh + \bg)^\top H^+ (\bh + \bg), \quad F^{(2)}_{00} = \frac{1}{2 \beta} \log \left[ \left(  \frac{\beta h_0^2}{2\pi} \right)^{dN} \det H \right] - \half (\bh - \bg)^\top H^+ (\bh - \bg)
\eeq
giving 
\beq
\D F_{00} = F^{(2)}_{00} - F^{(1)}_{00} = 2 \bh^\top H^+ \bg
\eeq 
Now we solve for $\D F$. Since the ligand binding applies a force, the curvature does not change, and the binding free energies only depend on the change in the energy minimum. Because we assume a symmetric protein, we can write the free energies in terms of either binding site. Here we choose site $a$.
\beq
\D F^{(1)} = F^{(1)}_{10} - F^{(1)}_{00} = -\frac{1}{2} f_a^2 \bu_a^\top H^+ \bu_a - f_a \bu_a^\top H^+ \bh - f_a \bu_a^\top H^+ \bg 
\eeq
and instead
\beq
\D F^{(2)} = F^{(2)}_{10} - F^{(2)}_{00} = -\frac{1}{2} f_a^2\bu_a^\top H^+ \bu_a - f_a\bu_a^\top H^+\bh + f_a \bu_a^\top H^+ \bg.
\eeq
Finally,
\beq
\D F = \D F^{(2)} - \D F^{(1)} = 2  f_a \bu_a^\top H^+ \bg.
\eeq

\clearpage

\subsection*{Two physical principles enable allostery: Arbitrary \texorpdfstring{$V_a, V_b$}{Va, Vb}}

In the main text, ligand potentials were assumed to act along a single collective mode,
$V_a(\br_a)=g_a(\bu_a^\top\br)$ and $V_b(\br_b)=g_b(\bu_b^\top\br)$.
Here we relax this assumption and allow arbitrary binding-site potentials $V_a(\br_a)$ and
$V_b(\br_b)$, and show that the same conclusions hold.

We partition the degrees of freedom as $\br=(\br_a,\br_r,\br_b)$, where $\br_a$ and $\br_b$
describe binding sites $a$ and $b$, respectively. The four ligation states are then
\beq
U_{\s_a\s_b}(\br)=U_p(\br)+\s_a V_a(\br_a)+\s_b V_b(\br_b).
\eeq
The corresponding partition functions can be written by isolating the binding-site coordinates,
\begin{align}
Z_{\s_a\s_b}
&=\int d\bx_a\,e^{-\s_a\beta V_a(\bx_a)}
   \int d\bx_b\,e^{-\s_b\beta V_b(\bx_b)}\,\tilde Z(\bx_a,\bx_b),
\end{align}
with
\beq
\tilde Z(\bx_a,\bx_b)=\int d\br\,
\delta(\br_a-\bx_a)\delta(\br_b-\bx_b)\,e^{-\beta U_p(\br)}.
\eeq
Using the Fourier representation of the delta functions and approximating $U_p$ by the multi-state quadratic free energy landscape~\eqref{eq:multistateLandscape}, we obtain
\beq
\tilde Z(\bx_a,\bx_b)
=\sum_{i=1}^n C_i
\exp\!\left[
-\frac{\beta}{2}
(\bar{\bx}-\bar{\bR}_i)^\top \bar{G}_i^{-1}(\bar{\bx}-\bar{\bR}_i)
-\beta\epsilon_i
\right],
\eeq
where $\bar{\bx}=(\bx_a,\bx_b)$, $\bar{\bR}_i=(\bR_i^a,\bR_i^b)$, $H^+_{i,ab}$ denotes the block of $H_i^+$ with rows in site $a$ and columns in site $b$, and
\beq
\bar{G}_i=
\begin{pmatrix}
H^+_{i,aa} & H^+_{i,ab} \\
(H^+_{i,ab})^\top & H^+_{i,bb}
\end{pmatrix}.
\eeq
The constants $C_i$ collect normalization factors that do not affect separability. The resulting free energy surface is
\beq \label{eq:arbitraryVaVb_freeEnergySurf}
\tilde F(\bx_a,\bx_b)
=-\beta^{-1}\log\!\left[
\sum_{i=1}^n
C_i
e^{-\frac{\beta}{2}
(\bar{\bx}-\bar{\bR}_i)^\top \bar{G}_i^{-1}(\bar{\bx}-\bar{\bR}_i)
-\beta\epsilon_i}
\right].
\eeq
For a single state ($n=1$), this reduces, up to an additive constant, to
\beq
\tilde F(\bx_a,\bx_b)
=\half(\bar{\bx}-\bar{\bR})^\top \bar{G}^{-1}(\bar{\bx}-\bar{\bR}).
\eeq
The free energy surface is separable if and only if the off-diagonal block of $\bar{G}^{-1}$
vanishes. For a symmetric $2\times2$ block matrix,
\beq
\bar{G}^{-1}_{ab}=0
\quad\Longleftrightarrow\quad
H^+_{ab}=0,
\eeq
assuming $H^+_{aa}$ and $H^+_{bb}$ are finite and invertible. Thus, even for arbitrary binding-site potentials $V_a$ and $V_b$, cooperativity in a single
state requires correlations between binding-site degrees of freedom encoded in $H^+_{ab}$. When there are two or more states, \eqref{eq:arbitraryVaVb_freeEnergySurf} need not be separable even when $H^+_{ab} = 0$. Just as in the simpler treatment of 1d ligand potentials, this multi-state non-separability is the basis for the MWC principle.

\subsection*{Negative cooperativity is possible in a symmetric dimer operating purely by MWC}
We consider a symmetric homodimer with two identical ligand-binding sites, labeled $a$ and $b$. The full conformation is $\br$, and $P$ is the linear swap operator that exchanges the two subunits (and hence $a\leftrightarrow b$), with $P^2 = I$ and $|\det P|=1$.  We assume the ligand potentials at the two sites are the same, $g_a = g_b = g$. We assume protein symmetry $U_p(\br)=U_p(P\br)$ and binding site symmetry $\bu_b = P \bu_a$.
These assumptions imply the correct transformation of ligation states under swap:
\beq
U_{00}(\br)=U_{00}(P\br),\qquad U_{11}(\br)=U_{11}(P\br),\qquad
U_{10}(\br)=U_{01}(P\br).
\eeq
We take the multi-basin model \eqref{eq:multistateLandscape} with $n$ basins. The partition  function in the unbound condition can be expressed as the sum of weights from each basin.
\beq
Z_{00}=\sum_{i=1}^n w_i,\quad \mathrm{with} \quad
w_i\equiv\int d\br
\exp\left[-\b\left(\half(\br-\bR_i)^\top H_i(\br-\bR_i) + \e_i\right) \right].
\label{eq:wi_def}
\eeq
The partition functions in the three bound conditions can be similarly expressed,
\beq
Z_{10}=\sum_{i=1}^n w_i Q^{a}_i,\qquad
Z_{01}=\sum_{i=1}^n w_i Q^{b}_i,\qquad
Z_{11}=\sum_{i=1}^n w_i Q^{ab}_i.
\eeq
with
\beq
Q^{a}_i \equiv \Big\langle e^{-\b g(\bu_a^\top\br)}\Big\rangle_i, \qquad
Q^{b}_i \equiv \Big\langle e^{-\b g(\bu_b^\top\br)}\Big\rangle_i, \qquad
Q^{ab}_i \equiv \Big\langle e^{-\b[g(\bu_a^\top\br)+g(\bu_b^\top\br)]}\Big\rangle_i.
\eeq
Here $\langle \cdot \rangle_i$ denotes an average over configurations within basin $i$ in the unbound condition. Explicitly, for any function $f(\br)$ we define
\beq
\langle f(\br)\rangle_i \;\equiv\;
\frac{\int d\br\; f(\br)\,
\exp\!\left[-\b\left(\half(\br-\bR_i)^\top H_i(\br-\bR_i) + \e_i\right) \right]}
{\int d\br\;\exp\!\left[-\b\left(\half(\br-\bR_i)^\top H_i(\br-\bR_i) + \e_i\right) \right]}.
\label{eq:basin_average_def}
\eeq
Because of symmetry,
\beq
Z_{10}=\sum_i w_iQ^{a}_i=\sum_i w_iQ^{b}_i=Z_{01}.
\eeq
We now impose the pure MWC condition of no intra-basin soft-mode coupling between the two sites, $\bu_a^\top H_i^{+}\bu_b = 0 \ \forall \ i$ which implies $\bu_a^\top\br$ and $\bu_b^\top\br$ are independent within each basin $i$. Therefore the two-site factor factorizes, $Q^{ab}_i=Q^{a}_i Q^{b}_i$, giving
\beq
Z_{11}=\sum_i w_i Q^{ab}_i=\sum_i w_i Q^{a}_i Q^{b}_i.
\eeq
Using $Z_{10}=Z_{01}$ and $Z_{11}=\sum_i w_iQ^{a}_iQ^{b}_i$, we obtain
\beq
\D\D F
= -\b^{-1}\log\!\Big(\frac{Z_{10}^2}{Z_{00}Z_{11}}\Big)
= -\b^{-1}\log\!\left[
\frac{\left(\sum_{i=1}^n w_iQ^{a}_i\right)^2}
{\left(\sum_{i=1}^n w_i\right)\left(\sum_{i=1}^n w_iQ^{a}_iQ^{b}_i\right)}
\right].
\label{eq:ddF_multibasin}
\eeq
Because $Q^{a}_i$ and $Q^{b}_i$ can vary across basins, the ratio in \eqref{eq:ddF_multibasin} is not constrained to be greater than or less than $1$, and therefore $\D\D F$ can be negative even though each basin satisfies pure MWC ($Q^{ab}_i=Q^{a}_iQ^{b}_i$).

\clearpage

\subsection*{Numerical parameters for figures}
Figures~\ref{fig:linearModels}--\ref{fig:2d_freeEnergy_landscape} use specific values for $H, H_{\rm dis}, \bu_a, \bu_b, \bg, \bh$ and other parameters. For all figures, we use systems of 5 dimensions and choose diagonal Hessians so that the entries of $\bu_a$ and $\bu_b$ clearly describe the effect on the normal modes. For all figures $\beta = 1$. The function $\rm{diag}$ takes a vector and returns a diagonal matrix with the vector's entries on the diagonal, and $\rm{normalize}(\bv) = \bv / \Vert \bv \Vert$. \\

\noindent \textbf{Parameters for Fig.~\ref{fig:linearModels}} \\
$H = \mathrm{diag}( ( 10, 0.5, 10, 10, 10 ) ),$
$\bu_a = \mathrm{normalize}( (0.0, 0.9, 0.1, 0, 0) ),$
$\bu_b = \mathrm{normalize}( (0.9, 0.1, 0.0, 0.1, 0) ).$\\

\noindent \textbf{Parameters for Fig.~\ref{fig:mechanisms} Left Half} \\
$H = \mathrm{diag} ( ( 10, \lambda, 10, 10, 10 ) )$,
$\bu_a = \mathrm{normalize}( (0.9, 0.1, 0.1, 0, 0) )$,
$\bu_b = \mathrm{normalize}( (0.9, 0.1, 0.0, 0.1, 0) )$,
$\bg = (g, 0,0,0,0)$,
$f_a = f_b = 1$,
$\bh = -(f_a \bu_a + f_b \bu_b) / 2$. \\

\noindent \textbf{Parameters for Fig.~\ref{fig:mechanisms} Right Half}\\
$H = \mathrm{diag} ( ( 10, \lambda, 10, 10, 10 ) )$,
$H_{\rm dis} = \mathrm{diag} ( ( 0.01, 0.01, 0.01, 0.01, 0.01 ) )$,
$\bu_a = \mathrm{normalize}( (0.0, 0.5, 0.5, 0, 0) )$, \\
$\bu_b = \mathrm{normalize}( (0.0, 0.5, 0.0, 0.5, 0) )$,
$k_a = k_b = 1$.\\

\noindent \textbf{Parameters for Fig.~\ref{fig:2d_freeEnergy_landscape} A }\\
$H = \mathrm{diag} ( ( 1, 0.1, 1, 1, 1 ) )$,
$H_{\rm dis} = \mathrm{diag} ( ( 0.01, 0.01, 0.01, 0.01, 0.01 ) )$,
$\bu_a = \mathrm{normalize}( (0.0, 0.5, 0.5, 0, 0) )$,\\
$\bu_b = \mathrm{normalize}( (0.0, 0.5, 0.0, 0.5, 0) )$,
$f_a = f_b = 0.5$,
$k_a = k_b = 0$,
$\e_{\rm dis} = 30$,
$\bh = -(f_a \bu_a + f_b \bu_b) / 2$,
$\bg = (0,0,0,0,0)$.\\

\noindent \textbf{Parameters for Fig.~\ref{fig:2d_freeEnergy_landscape} B }\\
$H = \mathrm{diag} ( ( 1, 0.08, 1, 1, 1 ) )$,
$H_{\rm dis} = \mathrm{diag} ( ( 0.01, 0.01, 0.01, 0.01, 0.01 ) )$,
$\bu_a = \mathrm{normalize}( (0.0, 0.5, 0.5, 0, 0) )$,\\
$\bu_b = \mathrm{normalize}( (0.0, 0.5, 0.0, 0.5, 0) )$,
$f_a = f_b = 0$,
$k_a = k_b = 0.5$,
$\e_{\rm dis} = 30$,
$\bh = (0,0,0,0,0)$,
$\bg = (0,0,0,0,0)$.\\

\noindent \textbf{Parameters for Fig.~\ref{fig:2d_freeEnergy_landscape} C }\\
$H = \mathrm{diag} ( ( 1, 1, 1, 1, 1 ) )$,
$H_{\rm dis} = \mathrm{diag} ( ( 0.01, 0.01, 0.01, 0.01, 0.01 ) )$,
$\bu_a = \mathrm{normalize}( (0.5, 0.0, 0.5, 0, 0) )$,\\
$\bu_b = \mathrm{normalize}( (0.5, 0.0, 0.0, 0.5, 0) )$,
$f_a = f_b = 0.5$,
$k_a = k_b = 0$,
$\e_{\rm dis} = 30$,
$\bh = -(f_a \bu_a + f_b \bu_b) / 2$,
$\bg = (5,0,0,0,0)$.\\

\noindent \textbf{Parameters for Fig.~\ref{fig:2d_freeEnergy_landscape} D }\\
$H = \mathrm{diag} ( ( 1, 1, 1, 1, 1 ) )$,
$H_{\rm dis} = \mathrm{diag} ( ( 0.01, 0.01, 0.01, 0.01, 0.01 ) )$,
$\bu_a = \mathrm{normalize}( (0, 0, 1, 0, 0) )$,\\
$\bu_b = \mathrm{normalize}( (0, 0, 0, 1, 0) )$,
$f_a = f_b = 0$,
$k_a = k_b = 0.5$,
$\e_{\rm dis} = 12$,
$\bh = (0,0,0,0,0)$,
$\bg = (0,0,0,0,0)$.\\

\noindent \textbf{Parameters for Fig.~\ref{fig:neg_coop} A-C }\\
$H = \mathrm{diag} ( ( 10, \lambda, 10, 10, 10 ) )$,
$\bu_a = \mathrm{normalize}( (0.9, 0.1, 0.1, 0, 0) )$,
$\bu_b = \mathrm{normalize}( (0.9, 0.1, 0.0, 0.1, 0) )$,
$\bg = (g, 0,0,0,0)$,
$f_a = 1, f_b = -1$,
$\bh = -(f_a \bu_a + f_b \bu_b) / 2$. \\

\noindent \textbf{Parameters for Fig.~\ref{fig:neg_coop} D-F}\\
$\bu_a = \mathrm{normalize}( (0.0, 0.5, 0.5, 0, 0) )$,
$\bu_b = \mathrm{normalize}( (0.0, 0.5, 0.0, 0.5, 0) )$,
$k_a = 1, k_b = -1$.
$H = \mathrm{diag} ( ( 10, \lambda, 10, 10, 10 ) ) - k_b\bu_b\bu_b^\top$,
$H_{\rm dis} = \mathrm{diag} ( ( 0.01, 0.01, 0.01, 0.01, 0.01 ) ) - k_b\bu_b\bu_b^\top$. These additions to the Hessians are done to keep the matrices positive semidefinite upon binding ligand b. Under these values the most flexible binding condition is $\s_a,\s_b = 0, 1$, as seen in panel F.

\clearpage
\subsection*{Supplementary Figures}
\renewcommand{\thefigure}{S\arabic{figure}}
\setcounter{figure}{0}

\begin{figure}[!h]
\centering
\includegraphics[width=0.65\linewidth]{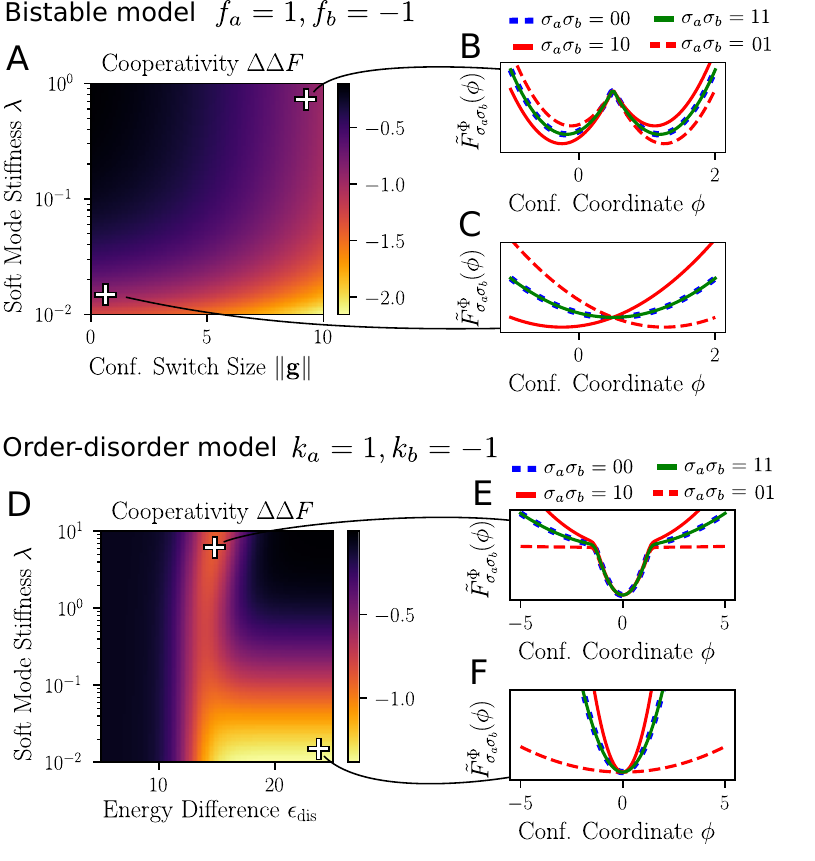}
\caption{Negative cooperativity in different cases.
Panels A–C are generated from a bistable landscape with bistability strength $\Vert \bg \Vert$ and a mode that couples the binding sites with stiffness $\lambda$. Ligand binding is modeled through applied forces, with $f_a = 1$ and $f_b = -1$, which push the system in opposing conformational directions. The full system is described by $U_{\s_a \s_b} = U_{\rm bi} - \s_a f_a \bu_a^\top \br - \s_b f_b \bu_b^\top \br$.
(A) The cooperativity $\D\D F$ takes negative values across the $\lambda$–$\Vert \bg \Vert$ parameter space.
(B) The free energy along the conformational-change coordinate, $\tilde{F}^{\Phi}_{\s_a \s_b}(\phi)$, for the system located at the upper-right $+$ mark in A, shown for the four ligation conditions. Because the ligand forces have opposing effects, there is minimal conformational change between the unbound and doubly bound conditions; therefore, to define $\tilde{F}^{\Phi}_{\s_a \s_b}(\phi)$ we use the conformational change between the two singly bound conditions, $\D \bR = \langle \br \rangle_{10} - \langle \br \rangle_{01}$. This regime corresponds to the conformational-switch case.
(C) $\tilde{F}^{\Phi}_{\s_a \s_b}(\phi)$ for the system at the bottom-left $+$ mark in A, again using $\D \bR = \langle \br \rangle_{10} - \langle \br \rangle_{01}$. This regime corresponds to the induced fit case.
Panels D–F are generated from the order–disorder landscape, in which the ordered state has a mode that couples the binding sites with stiffness $\lambda$, and the ordered and disordered states are separated in energy by $\e_{\rm dis}$. Ligand binding is modeled as applied stiffness with opposing effects: binding at site $a$ increases the stiffness ($k_a = 1$) and binding at site $b$ decreases the stiffness ($k_b = -1$). The full system is described by $U_{\s_a \s_b} = F_{\rm q,d} + \s_a \tfrac{k_a}{2} (\bu_a^\top \br)^2 + \s_b \tfrac{k_b}{2} (\bu_b^\top \br)^2$.
(D) The cooperativity $\D\D F$ takes negative values across the $\lambda$–$\e_{\rm dis}$ parameter space.
(E) The free energy along a conformational coordinate, $\tilde{F}^{\Phi}_{\s_a \s_b}(\phi)$, for the system located at the top-center $+$ mark in D, shown for the four ligation conditions. This regime corresponds to the order–disorder case.
(F) $\tilde{F}^{\Phi}_{\s_a \s_b}(\phi)$ for the system at the bottom-right $+$ mark in D, shown for the four ligation conditions. This regime corresponds to the dynamic case.
Because there is no net conformational change upon binding in E and F, we take $\D \bR$ in \eqref{eq:cv_Phi} to be the soft mode. Numerical values for the plots are listed above under ``Numerical parameters for figures.''}
\label{fig:neg_coop}
\end{figure}

\end{document}